\documentclass[fleqn,usenatbib]{mnras}
\usepackage{newtxtext,newtxmath}

\usepackage[T1]{fontenc}
\usepackage{ae,aecompl}
\usepackage{graphicx}	
\usepackage{amsmath}	
\usepackage{amssymb}	
\usepackage[english]{babel}
\usepackage[utf8]{inputenc}
\newcommand{\jvec}{\mathbf{j}}
\newcommand{\evec}{\mathbf{e}}

\newcommand{\uvec}{\hat{\mathbf{u}}}
\newcommand{\nvec}{\hat{\mathbf{n}}}

\title[Short title, max. 45 characters]{MNRAS \LaTeXe\ template -- title goes here}

\author[Pu \& Lai]{
Bonan Pu\thanks{E-mail: bonanpu@astro.cornell.edu (BP)} and Dong Lai
\\
Cornell Center for Astrophysics and Planetary Science, Department of Astronomy, 
Cornell University, Ithaca, NY 14853, USA\\
}

\date{Accepted XXX. Received YYY; in original form ZZZ}

\pubyear{2015}

\begin{document}

\title{Eccentricities and Inclinations of Multi-Planet Systems with External Perturbers}
\date{\today}
\maketitle

\begin{abstract} 
Compact multi-planet systems containing super-Earths or sub-Neptunes,
commonly found around solar-type stars, may be surrounded by external
giant planet or stellar companions, which can shape the architechture
and observability of the inner systems.  We present a comprehensive
study on the evolution of the inner planetary system subject to the
gravitational influence of an eccentric, misaligned outer perturber.
Analytic results are derived for the inner planet eccentricities
($e_i$) and mutual inclination ($\theta_{12}$) of the ``2-planet +
perturber'' system, calibrated with numerical secular and N-body
integrations, as a function of the perturber mass $m_p$, semi-major
axis $a_p$ and inclination angle $\theta_p$. We find that the dynamics
of the inner system is determined by the dimensionless parameter
$\epsilon_{12}$, given by the ratio between the differential
precession rate driven by the perturber and the mutual precession rate
of the inner planets.  Loosely packed systems  (corresponding to $\epsilon_{12} \gg 1$) are more susceptible to eccentricity/inclination excitations by the perturber than tightly packed inner systems (with $\epsilon_{12} \ll 1$) (or singletons),
although resonance may occur around $\epsilon_{12}\sim 1$,
leading to large $e_i$ and $\theta_{12}$. Dynamical instability may set in for inner planet systems with large excited eccentricities and mutual inclinations. We present a formalism to extend our analytical results to general
inner systems with $N>2$ planets and apply our results to constrain
possible external companions to the Kepler-11 system. Eccentricity and inclination excitation by external companions may help explain the observational trend that systems with fewer transiting planets are dynamically hotter than those with more transiting planets.
\end{abstract}



\section{Introduction}

Recent advances in radial velocity and transit
surveys have led to a large increase in the number of detected
multi-planet systems, indicating that such systems are common in the Galaxy.
Of particular interest are the compact multi-planet systems discovered by NASA's
{\it Kepler} mission \citep{Mullally2015,Burke2015,Morton2016}. 
Such systems generally feature multiple super-Earths or sub-Neptunes (with radii
1.2-3$R_\oplus$) with periods inwards of 200 days. The orbital configurations of these systems 
suggest that they are generally ``dynamically cold'', with 
eccentricities $e \sim 0.02$ \citep{Lithwick2012, VanEylen2015,
  Xie2016, Hadden2017}, orbital mutual inclinations $\theta \sim
2^\circ$ \citep{Lissauer2011,
  Fang2012,Figueira2012,Johansen2012,Tremaine2012,Fabrycky2014}, and
orbital spacings close to the limit of stability \citep{Fang2013,Pu2015,Volk2015}.

What account for the origins of these orbital properties? One
suggestion is that these planets have orbital configurations that
reflect their formation environments \citep{Hansen2013,Moriarty2016} -
with the ``dynamically cold'' population being an indicator of having
formed in highly dissipative environments or vice-versa. Indeed, there
is evidence for a correlation between planet compositions and orbits:
planets with more gaseous envelops tend to have dynamcially ``colder'' orbits,
consistent with a gaseous formation environment \citep{Dawson2016}.
 
Another possibility is that the orbital properties of such planets are
driven by external perturbers, of either the planetary or stellar
variety.  Stellar companions to numerous Kepler systems have been
detected in a wide range of separations \citep{Baranec2016}. There is
evidence for a reduced occurrence of stellar companions to stars hosting multiple transiting planets: \cite{Wang2015b} found that $5\pm 5\%$ of Kepler multi's have stellar companions at separation 1-100~au in comparison to the rate of 21 \% for field stars, indicating that stellar companions may be disruptive to the formation
or stability of cold multi-planet systems. Several proto-planetary disks have been observed to be misaligned with their binary companions \citep{Jensen2014, Brinch2016}, therefore misaligned binary companions to planetary systems may be common.

On the other hand, a number of long-period giant planet companions to
Kepler compact systems have been found using the transit method
\citep{Schmitt2014, Uehara2016} and the RV method (e.g.,
Kepler-48, Kepler-56, Kepler-68, Kepler-90, Kepler-454). A few
non-Kepler ``inner compact planets + giant companion'' systems have
also been discovered (e.g., GJ 832, WASP-47). \cite{Bryan2016} reported that about 50\% of one and two-planet systems discovered by RV have companions in the 1-20 $M_J$ and 5-20 au range.  These results
indicate that external ($\gtrsim$1 au) giant planet companions are
common around hot/warm ($\lesssim$1 au) planets, and may significantly
shape the architecture of the inner planetary systems.

The outer stellar or planetary companions can influence their inner
systems in a variety of ways, changing their orbital properties and
even making them dynamically unstable. The most common effects arise
from secular gravitational perturbation.  In general, a distant
stellar companion may be on an inclined orbit relative to the inner
planetary system because of its misaligned orbital angular momentum at
birth.  A giant planet may also have an inclined and eccentric orbit,
as a result of strong scatterings between multiple giants \citep{Juric2008,Chatterjee2008}.  Such misaligned external
perturbers can induce mutual inclination oscillations amongst the
inner planets, making the inner system dynamically hotter or even
unstable \citep{Lai2017, Hansen2017, Becker2017, Read2017,
  JontofHutter2017}. While mild inclination oscillations tend to preserve the
integrity of the inner systems, they can nevertheless disrupt the
co-transiting geometry of the inner planets and thereby reduce the
number of transiting planets \citep[e.g.][]{Brakensiek2016}.  In \cite{Lai2017}, we have derived approximate analytic expressions,
calibrated with numerical calculations, for the mutual inclination
excitations for various planetary systems and perturber properties; we
have also identified a secular resonance mechanism that can generate
large mutual inclinations even for nearly co-planar perturbers. 

A more subtle effect of external perturbers on inner planetary systems
is ``evection resonance'' \citep{Touma2015,Xu2016}.
This occurs when the apsidal precession
frequencies of the inner planets, driven by mutual gravitational
interactions, match the orbital frequency of the external companion.
Resonant eccentricty excitation in the inner planets may lead to diruption 
of the system under some conditions.

A more ``violent'' scenario involves two or more giant planets in an 
unstable configuration, leading to multiple close encounters and
ejections/collisions of planets that finally end when a stable configuration is reached 
- usually with a single giant planet remaining \citep{Chatterjee2008,Juric2008}. 
This ``outer violence'' can excite the eccentricities and inclinations of the inner
systems, often to their demise - although the end result is highly
variable, depending on the initial separations between the giants and
inner planets \citep{Matsumura2013, Carerra2016, Huang2017, Gratia2017, Mustill2017}.
One can view this ``violent phase'' as the precursor of the
``secular phase'' mentioned above:
Two or three giant planets first form in nearly circular, co-planar configurations ; the planets then undergo strong 
scatterings. This is a natural mechanism of producing a misaligned
giant planet that can induce further secular eccentricity/inclination excitations
in the inner planetary system. Indeed, there is an interplay between the ``violent'' 
and ``secular'' phases of such planetary systems (Pu \& Lai 2018, in prep); the results of this paper serves as a baseline for our forthcoming exploration on the eccentricity and inclination excitation during the ``dynamical'' phase.

In this paper, we study the evolution of multi-planet inner
systems with a single eccentric, misaligned outer companion (star or
giant planet). We develop tools based on secular perturbation theory,
calibrated with N-body simulations, to predict the outcomes of inner
planet orbital properties based on the perturber's orbits. We extend our previous
work \citep{Lai2017} to treat the combined excitations of eccentricities and
mutual inclinations. A major goal of our paper 
is to derive approximate analytic expressions and scaling formulae
that can be adapted to various planetary and perturber parameters.

The framework of this paper is as follows. In Section \ref{sec:sec2}, we apply linear Laplace-Lagrange theory to derive analytic expressions for the
evolution of eccentricities and inclinations of planets in a ``2-planets + perturber'' system. In Section \ref{sec:sec3} (see also Appendix \ref{sec:appendix}), we develop an approximate procedure to extend our analytic results to the mildly non-linear regime (with the perturber's inclination and eccentricity satisfying $\theta_p, e_p \lesssim 0.4$); within this regime, our analytic results agree robustly with numerical results
based on integration of secular equations. In Section \ref{sec:sec4} we compare
our theoretical results based on secular theory to N-body
simulations. In Section \ref{sec:sec5} we extend our results
to inner systems with $N > 2$ planets, and a prescription for evaluating the planet RMS eccentricities and mutual inclinations in a ``N-planets + perturber'' system is given in Appendix \ref{sec:appendix2}. In Section \ref{sec:sec6} we illustrate our results by applying them to the Kepler-11 system. We summarize our key findings in Section \ref{sec:sec7}.

\section{Eccentrcity and Inclination Excitation in Linearized Secular Theory}
\label{sec:sec2}

Consider an inner system of planets with masses $m_i$ ($i =
1,2,3...,N$) and semi-major axes $a_i$ ($a_1 < a_2 < ... <
a_N$). These inner planets are initially coplanar and have circular
orbits, and are accompanied by a giant planet (or stellar) perturber with
$m_p \gg m_i$, semi-major axis $a_p \gg a_i$, inclination angle
$\theta_p$ and eccentricity $e_p$.  How do the eccentricities and
mutual inclinations of the inner planet system evolve?


In this Section we consider the regime where all eccentricities and
inclinations are small ($e_p, e_i \ll 1$ and $\theta_p, \theta_i \ll 1$). 
In this regime, the evolutions of the eccentricities $e_i$
and inclinations $\theta_i$ decouple, and are governed by the standard
linearized Laplace-Lagrange planetary equations \citep{MurrayDermott}. We
present several analytical results that will be useful for the more
general cases where $e_p$ and $\theta_p$ are more modest.

We define the dimensionless eccentricity vector $\mathbf{e}_j$ and
dimensionless angular momentum vector $\jvec_j$ of the $j$-th planet as 
\begin{equation}
{\bf j}_j=\sqrt{1-e^2}{\bf \hat n}_j,\quad 
{\bf e}_j=e\,{\bf\hat u}_j
\label{eqn:evec_jvec}
\end{equation}
where ${\nvec}_j$ and ${\uvec}_j$ are unit vectors, and note that $\jvec_j \simeq {\bf \hat n}_j$ since $e_j \ll 1$. The evolution equations for $\evec_j$ and $\jvec_j$ (where $j = 1,2,3...N,p$, with the perturber included) are:
\begin{align}
\frac{d \mathbf{e}_j}{dt} &= - \sum_{k \neq j} \omega_{jk} (\mathbf{e}_j \times \jvec_k) - \sum_{k \neq j} \nu_{jk} (\jvec_j \times \evec_k), 
\label{eq:LL_e}\\
\frac{d \jvec_j}{dt} &= \sum_{k \neq j} \omega_{jk} (\jvec_j \times \jvec_k ).
\label{eq:LL_l}
\end{align}
The quantities $\omega_{jk}$ and $\nu_{jk}$ are the quadrupole and
octupole precession frequencies of the $j$-th planet due to the action
of the $k$-th planet, given by:
\begin{align}
\omega_{jk} = \frac{G m_j m_k a_{<}}{a_{>}^2 L_j } b^{(1)}_{3/2}(\alpha),
\label{eq:wjk} \\
\nu_{jk} = \frac{G m_j m_k a_{<} }{a_{>}^2 L_j} b^{(2)}_{3/2}(\alpha)
\label{eq:vjk}.
\end{align}
Here $a_< = {\rm min}(a_j,a_k)$, $a_> = {\rm max}(a_j,a_k)$, $\alpha = a_< / a_>$, 
$L_j \simeq m_j\sqrt{GM_*a_j}$ is the angular momentum of the $j$-th planet, and the $b^{(n)}_{3/2}(\alpha)$ 
are the Laplace coefficients defined by
\begin{equation}
b^{(n)}_{3/2}(\alpha) = \frac{1}{2\pi} \int_{0}^{\pi} \frac{\cos{(nt)}}{(\alpha^2 + 1 - 2\alpha \cos{t})^{3/2}} dt.
\end{equation}
Note that: 
\begin{equation}
\omega_{jk}L_j=\omega_{jk}L_k.
\end{equation}
For $\alpha \ll 1$, we have $b^{(1)}_{3/2}(\alpha) \simeq 3\alpha /4 +
43\alpha^3/32 + 525\alpha^5/256 $ and $b^{(2)}_{3/2}(\alpha) \simeq
15\alpha^2/16 + 105 \alpha^4/64$. Thus the 
ratio of the quadrupole and octupole frequencies is given by
\begin{equation}
\frac{\nu_{jk}}{\omega_{jk}} = \beta(\alpha)
\equiv \frac{b^{(2)}_{3/2}(\alpha)}{ b^{(1)}_{3/2}(\alpha)}
\simeq 5\alpha/4 - 5\alpha^3/32.
\label{eqn:beta12}
\end{equation}
It is usually sufficient to take $\beta(\alpha) \simeq 5\alpha/4$, as this is accurate to 
within $8\%$ for $\alpha \le 0.8$. Note that $\beta(\alpha) < 1$ for all $\alpha$; in the limit that $\alpha \rightarrow 1$, we have $\beta(\alpha) \rightarrow 1$ from below.


For convenience, we introduce the variables $\mathcal{I}$ and $\mathcal{E}$ as the
complex inclination and eccentricity:
\begin{align}
\mathcal{I} \equiv |\mathcal{I}|\, \exp ({i\Omega}), \label{eq:complex_i} \\
\mathcal{E} \equiv |\mathcal{E}|\, \exp ({i\varpi}),
\label{eq:complex_e}
\end{align}
where $\Omega$ and $\varpi$ are the longitude of the ascending note 
and the longitude of pericenter, respectively.

\subsection{``One Planet + Perturber'' System: Eccentricity}

\begin{figure}
\includegraphics[width=1.05\linewidth]{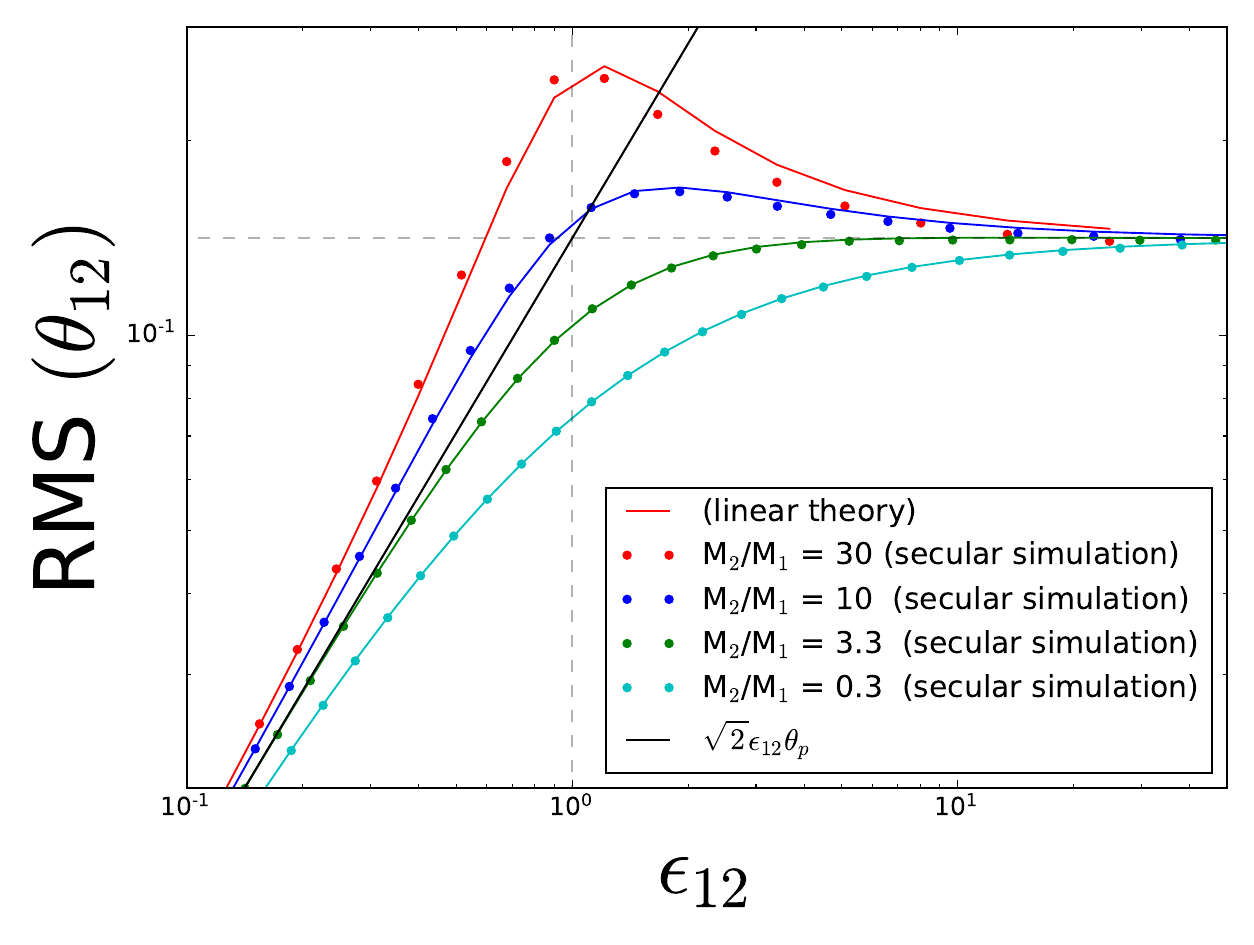}
\caption{RMS values of the mutual inclination between the inner planets
  in a fiducial two-planet system under the influence of a mis-aligned perturber.
  The perturber has initial eccentricity $e_p = 0$ and
  inclination $\theta_p = 0.1$ rad. Planet 1 has a
  fixed mass at $m_1 = M_{\oplus}$ while the mass of
  planet 2 varies from $0.3 M_{\oplus}$ to $30 M_{\oplus}$ and are
  represented by different colors. The perturber has a mass of $m_p =
  3 M_J$ and its semi-major axis is varied to produce different $\epsilon_{12}$, the
  coupling strength of the perturbation (Eq. \ref{eq:ep12}).
  The solid colored points represent the results of numerical integrations using
  secular equations, while the solid colored curves are calculated using linearized theory.
  The dashed vertical line corresponds to $\epsilon_{12} = 1$, where a resonance feature
  occurs. Note that the resonance feature becomes sharper and more
  pronounced as the mass ratio $m_2 / m_1$ increases. The horizontal
  dashed line corresponds to the weak-coupling limit
  $\mathrm{RMS}(\theta_{12}) \simeq \sqrt{2}\theta_p$, which holds when $\epsilon_{12} \gg 1$. The solid black line is the prediction from
secular theory, when the planets are strongly coupled ($\epsilon_{12} \ll 1$). }
  \label{fig:fig1}
\end{figure}

\begin{figure*}

\includegraphics[width=0.495\linewidth]{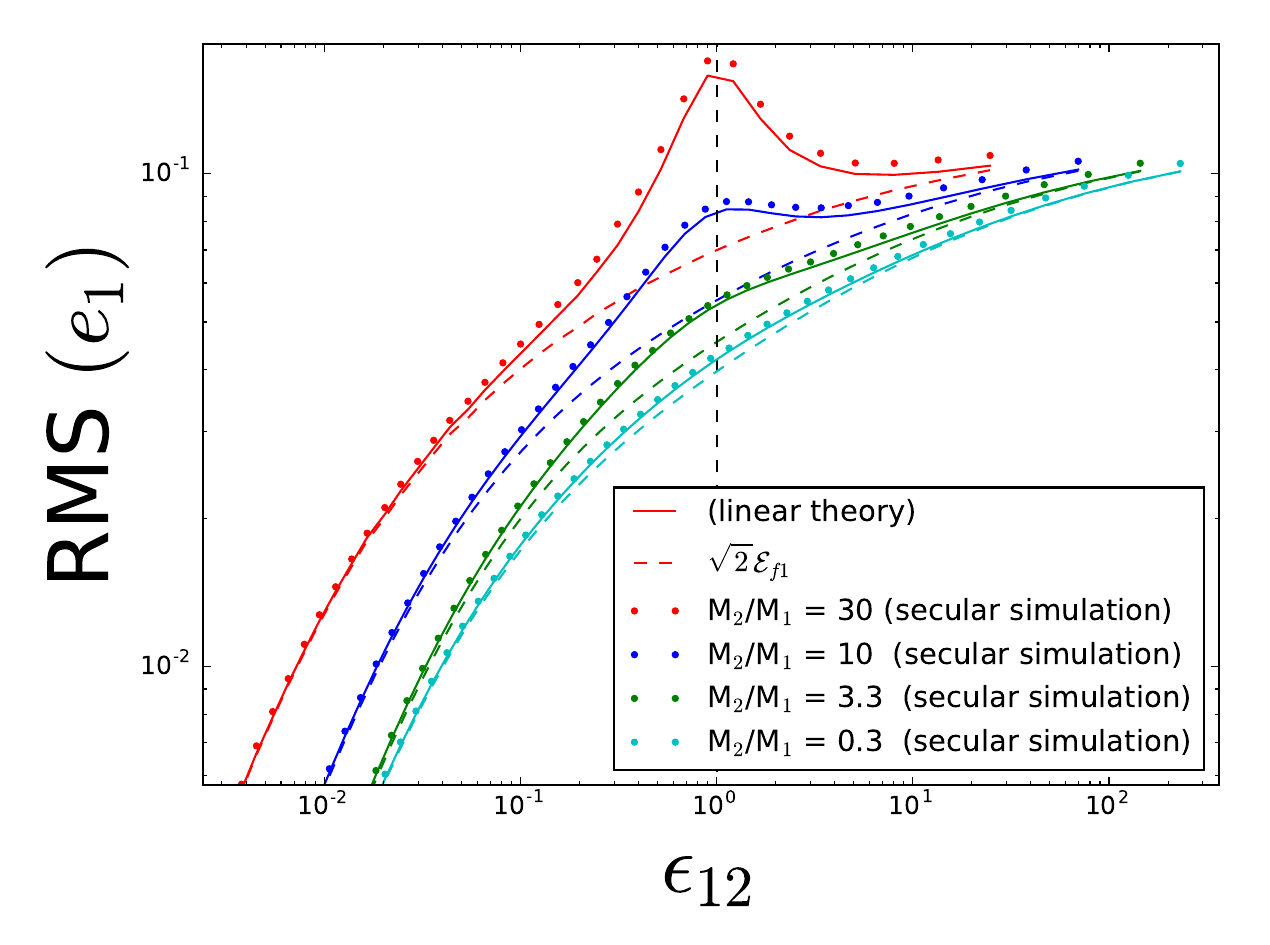}
\includegraphics[width=0.49\linewidth]{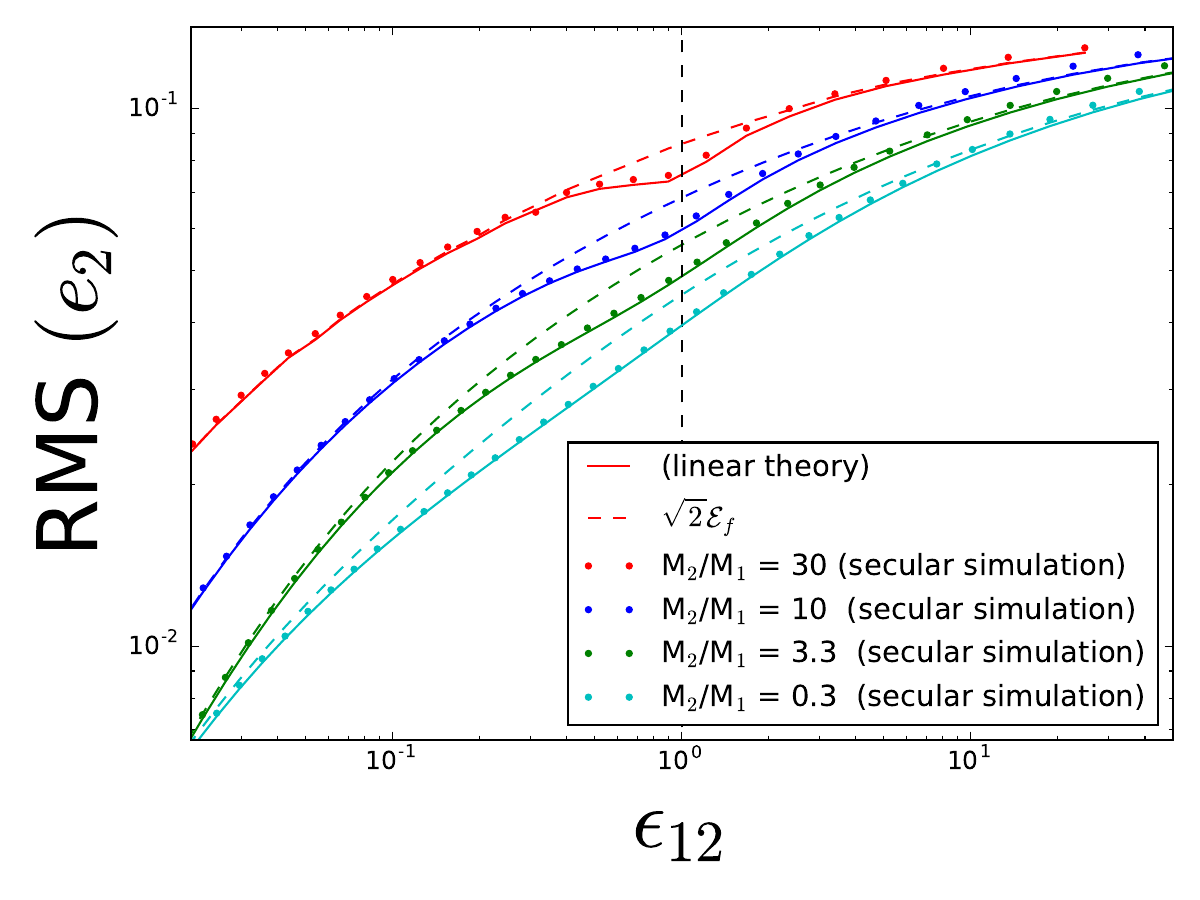}
\caption{RMS values of the eccentricities of the planets in a fiducial two-planet system under the influence         
  of an eccentric perturber. The perturber has initial eccentricity $e_p = 0.1$ and
  has an orbit co-planar with the inner system. The other parameters are the same as Fig. \ref{fig:fig1}.
  The solid colored points represent the results of numerical integrations using
  secular equations, while the solid colored curves are calculated using linearized theory;
  the dashed colored lines show the forced eccentricity (Eq. \ref{eqn:e_forced}).
  The dashed vertical line corresponds to $\epsilon_{12} = 1$.}
  \label{fig:fig2}
\end{figure*}

In the limit $m_p \gg m_1$, the eccentricity vector of the inner planet evolves in time as
\begin{equation}
\mathcal{E}_1(t) = \frac{\nu_{1p}}{\omega_{1p}} \mathcal{E}_p \left[1 - \exp({i\omega_{1p}t})\right]
\label{eqn:e1_t}
\end{equation}
assuming $\mathcal{E}_1(t = 0) = 0$. 
The root-mean-square (RMS) value of the eccentricity is therefore
\begin{equation}
\langle e_1^2 \rangle^{1/2} = \langle |\mathcal{E}_1|^2 \rangle^{1/2} 
= \sqrt{2}  \frac{\nu_{1p}}{\omega_{1p}} e_p \simeq \sqrt{2} \left(\frac{5}{4}\right) \left(\frac{a_1}{a_p}\right) e_p.
\label{eqn:e1_RMS}
\end{equation}
The maximal eccentricity reached by planet 1 is
\begin{equation}
(e_1)_{\rm max} =  2 \frac{\nu_{1p}}{\omega_{1p}} e_p \simeq \left(\frac{5}{2}\right) \left(\frac{a_1}{a_p}\right) e_p.
\label{eqn:e1_max}
\end{equation}

The above expressions assume $L_p \gg L_1$ and neglect the eccentricity evolution of the perturber. 
For finite $L_1/L_p$, the RMS value of $e_1$ can be generalized to
\begin{align}
\langle e_1^2 \rangle^{1/2} &= \frac{\sqrt{2} \nu_{1p} e_p}{\sqrt{(\omega_{1p} -
    \omega_{p1})^2 + 4 \nu_{1p} \nu_{p1}}} \nonumber\\
&\simeq \sqrt{2} \left(\frac{5}{4}\right) \left(\frac{a_1}{a_p}\right) e_p 
\left[\left(1-\frac{L_1}{L_p}\right)^2 + \frac{25}{4}\left(\frac{L_1}{L_p}\right)
  \left(\frac{a_1}{a_p}\right)^2 \right]^{-1/2}.
\end{align}

\subsection{``Two Planets + Perturber'' System: Mutual Inclination}

\begin{figure*}
\includegraphics[width=0.9\linewidth]{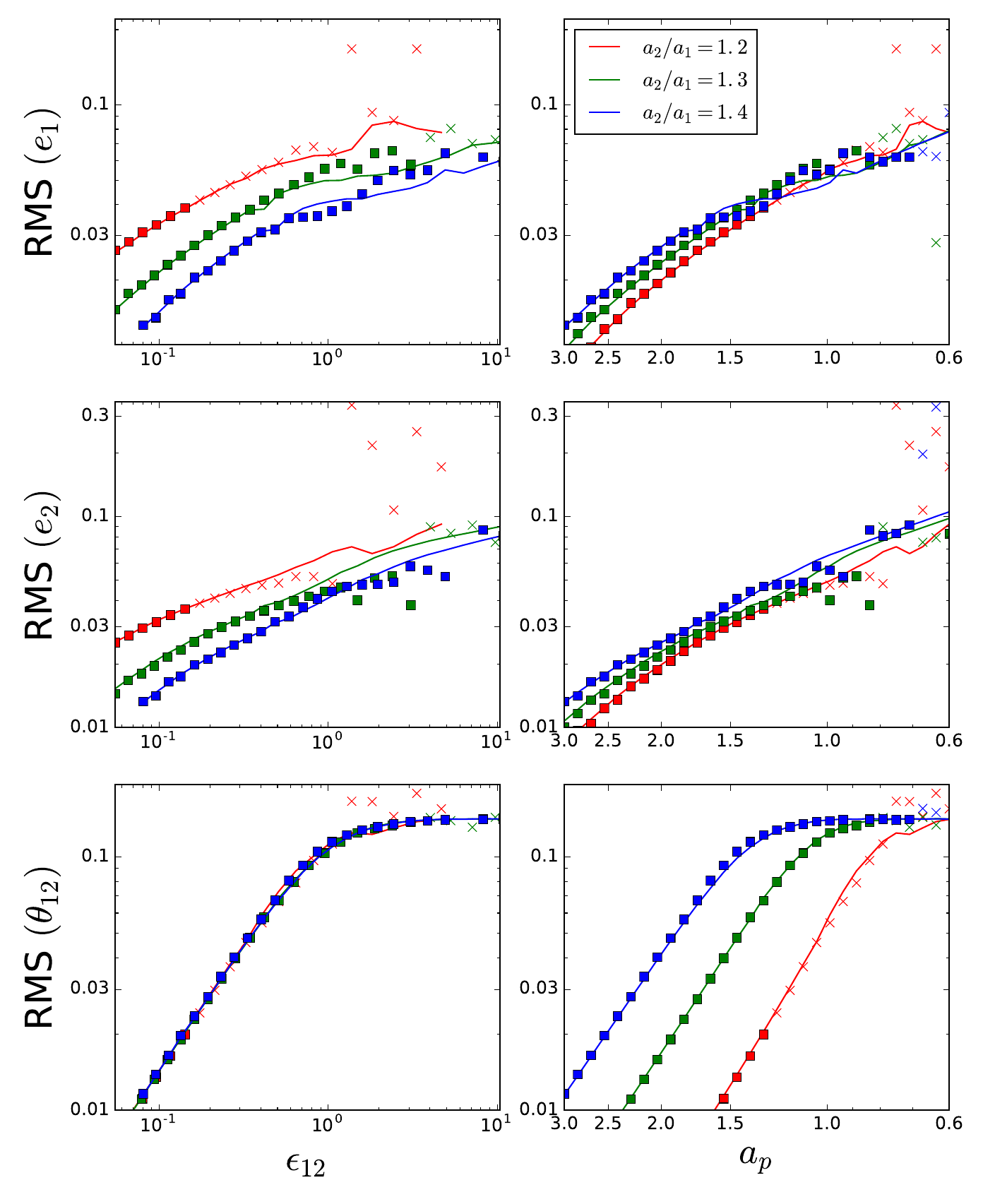}
\caption{RMS values of $e_1, e_2$ and $\theta_{12}$ (top to bottom) as a function of $\epsilon_{12}$ and $a_p$ for a 2-planet system with different spacings $a_2/a_1$. The solid curves are results of numerical integrations using the hybrid secular equations (Eqs. \ref{eq:A1} - \ref{eq:A4}) while the results of N-body simulations are marked with an ``$\times$'' if the system becomes gravitationally unstable with respect to orbit crossings within $10^5 $yr, otherwise they are marked with a filled square. The different colored lines represent different values of $a_2/a_1$, with red, green and blue being $a_2/a_1 = 1.2, 1.3, 1.4$ respectively. In each case, $a_1 = 0.3$ au, and $a_p$ is varied from 0.6 to 3.0 au. The planet masses are $m_1 = m_2 = 3M_{\oplus}$ and $m_p = 3M_J$. The inner planets are initially on circular and co-planar orbits while the perturber has $e_p = 0.05$ and $\theta_p = 0.1$. }
\label{fig:fig3}
\end{figure*}

In a previous paper \citep{Lai2017}, we have already studied the
secular evolution of the mutual inclination angle between the inner planets
in the presence of a misaligned outer companion. 
It was found that the evolution of the inner system depends
critically on the ratio of the differential quadrupole precession frequency driven by the
perturber and the mutual quadrupole precession frequency between the
inner planets, succintly described by the dimensionless parameter
$\epsilon_{12}$ (note that we adopt a change of notation from the
previous paper, $\Omega_{ip}$ is now $\omega_{ip}$):
\begin{align}
\epsilon_{12} &\equiv \frac{\omega_{2p}-\omega_{1p}}{\omega_{12}+\omega_{21}} \nonumber \\
&\approx \left(\frac{m_p}{10^3m_2}\right)\left(\frac{10a_2}{a_p}\right)^3
\left[\frac{3 a_1/a_2}{b_{3/2}^{(1)}(a_1/a_2)}\right] \frac{(a_2/a_1)^{3/2}-1}{1+(L_1/L_2)}.
\label{eq:ep12}
\end{align}
When $\epsilon_{12} \gg 1$, the perturber is
dominant and the inner planets behave as if they are indepedent of one
another; when $\epsilon_{12} \ll 1$, the inner planets are tightly
coupled and their angular momenta stay closely aligned, with 
a mutual inclination $\theta_{12} \sim \epsilon_{12}\theta_p$. 
In the regime where $\epsilon_{12} \sim 1$, a secular resonance
feature exists and it is possible for the inner planets to have
$\theta_{12} \gg \theta_p$.

The general expression for the mutual inclination, in the limit of $m_p \gg m_j$, is given by:
\begin{equation}
\langle \sin^2{\theta_{12}}\rangle^{1/2} = 2\theta_p \left( \frac{\omega_{2p} - \omega_{1p}}{\sqrt{(\omega_1 - \omega_2)^2 + 4\omega_{12}\omega_{21}}} \right),
\label{eq:theta12}
\end{equation}
where
\begin{align}
\omega_1=\omega_{12}+\omega_{1p},
\label{eq:w12_a} \\
\omega_2=\omega_{21}+\omega_{2p}.
\label{eq:w12_b}
\end{align}
It is clear that a resonance occurs when $\omega_{1} = \omega_{2}$
\footnote{Equation (\ref{eq:theta12}) is valid for $\theta_p\ll 1$. See \cite{Lai2017} for 
the result and the property of the resonance when $\theta_p$ is not restricted to small values.}.
At the resonance, we have
\begin{equation}
\langle \sin^2{\theta_{12}}\rangle^{1/2}_{\rm res} =\theta_p \sqrt{\frac{L_2}{L_1}}\left(1 - \frac{L_1}{L_2}\right).
\end{equation}

In the weak coupling limit ($\epsilon_{12} \ll 1$) the mutual inclination is given by
\begin{equation}
\langle \sin^2{\theta_{12}}\rangle^{1/2} \simeq \langle \theta_{12}^2 \rangle^{1/2} \simeq  \sqrt{2} \theta_p. 
\label{eq:theta_12_weak}
\end{equation}
In the above expression, we have neglected the back-reaction on the perturber by the inner planets. In the weak coupling regime, this feedback is of order $L_2/L_p$ and we find 
\begin{equation}
\langle \sin^2{\theta_{12}}\rangle^{1/2}  \simeq \langle \theta_{12}^2 \rangle^{1/2} \simeq \sqrt{2}\theta_p \left(1 + \frac{L_2}{L_p}\right)^{-1}.
\end{equation}
In the strong coupling limit ($\epsilon_{12}\ll 1$), the back-reaction is always negligible, and we have
\begin{equation}
\langle \sin^2{\theta_{12}}\rangle^{1/2} \simeq \langle \theta_{12}^2 \rangle^{1/2} \simeq \sqrt{2} \epsilon_{12}\theta_p.
\label{eq:theta_12_strong}
\end{equation}

The above results are summarized in Fig. \ref{fig:fig1}, in which we show the root-mean-square (RMS) mutual inclination between the inner planets in a fiducial two-planet system under the influence of an external misaligned companion. The linearized analytic results (solid curves) are shown to be in excellent agreement with secular numerical integrations (dots).

\subsection{``Two Planets + Perturber'' System: Eccentricity}

The eccentricty vectors of the two inner planets are governed by
\begin{align}
\frac{d}{dt} \begin{pmatrix} \mathcal{E}_1 \\ \mathcal{E}_2  \end{pmatrix} &= i \begin{pmatrix} \omega_{1} & -\nu_{12} \\ -\nu_{21} & \omega_{2} \end{pmatrix}  \begin{pmatrix} \mathcal{E}_1 \\ \mathcal{E}_2 \end{pmatrix} - i \begin{pmatrix} \nu_{1p} \\ \nu_{2p} \end{pmatrix}  \mathcal{E}_p  
\nonumber\\
&\equiv i \mathbf{A}  \begin{pmatrix} \mathcal{E}_1 \\ \mathcal{E}_2 \end{pmatrix} - i \mathbf{B}\, \mathcal{E}_p,
\label{eq:de12}
\end{align}
where $\omega_1$ and $\omega_2$ are given by Eqs. (\ref{eq:w12_a}) - (\ref{eq:w12_b}).

The homogeneous equation of (\ref{eq:de12}) (with $\mathcal{E}_p=0$) has two modes, 
with eigen-frequencies
\begin{align}
\lambda_{\pm} &= \frac{1}{2} \left(\omega_{1} + \omega_{2} \pm \gamma \right)\quad
{\rm with}~~ \gamma \equiv \sqrt{\Delta \omega^2 + 4 \nu_{12}\nu_{21}},
\label{eq:lambda_pm}
\end{align}
where $\Delta \omega \equiv (\omega_1- \omega_2)$ (the ``distance'' from the resonance).
Note $\gamma$ is at a minimum at the resonance ($\Delta \omega = 0$).
The corresponding eigenvectors are:
\begin{align}
\mathbf{v}_{\pm} &= \begin{pmatrix} \frac{\Delta \omega \pm \gamma}{ 2\nu_{21}} \\1 \end{pmatrix}. \label{eq:v_pm}
\end{align}

The forcing term in Eq. (\ref{eq:de12}) gives the inner planets a forced eccentricity
\begin{equation}
\begin{pmatrix} \mathcal{E}_{f1} \\ \mathcal{E}_{f2} \end{pmatrix} =  \mathbf{A}^{-1}\mathbf{B}
\,\mathcal{E}_p  =  \begin{pmatrix} F_{11} \\ F_{12} \end{pmatrix} e_p,
\label{eqn:e_forced}
\end{equation}
where 
\begin{align}
F_{11} &= \frac{\nu_{1p}\omega_2 + \nu_{12}\nu_{2p}} {\omega_1 \omega_2 - \nu_{12}\nu_{21} },\\
F_{12} &= \frac{\nu_{2p}\omega_1 + \nu_{21}\nu_{1p}} {\omega_1 \omega_2 - \nu_{12}\nu_{21} }.
\end{align}
If the inner planets are on 2 initially circular orbits, the general solution to equation (\ref{eq:de12}) is then
\begin{equation}
\begin{pmatrix} \mathcal{E}_1(t) \\ \mathcal{E}_2(t)  \end{pmatrix}
= \begin{pmatrix} \mathcal{E}_{f1} \\ \mathcal{E}_{f2}  \end{pmatrix}
+ c_+ \mathbf{v}_+ \exp{(i \lambda_+ t)} + c_- \mathbf{v}_- \exp{(i \lambda_- t)},
\label{eqn:gen_et}
\end{equation}
where the coefficients $c_\pm$ are determined by the initial conditions
\begin{align}
c_+ &=  +  \frac{\nu_{21}}{\gamma}\left(\frac{\mathcal{E}_{f2}(\Delta \omega + \gamma)}{2\nu_{21}}
                                              - \mathcal{E}_{f1}\right)  \label{eq:c_pm_1}                                            \\ 
c_- &=  -  \frac{\nu_{21}}{\gamma}\left(\frac{\mathcal{E}_{f2}(\Delta \omega - \gamma)}{2\nu_{21}}
                                              - \mathcal{E}_{f1}\right).
\label{eq:c_pm_2}                                            
\end{align} 
Note that $c_+ \mathbf{v_+} + c_- \mathbf{v_-} =
\mathbf{\mathcal{E}_f}$. One can verify that Eq. (\ref{eqn:gen_et})
is equivalent to
\begin{equation}
\begin{pmatrix} \mathcal{E}_1(t) \\ \mathcal{E}_2(t)  \end{pmatrix} =   -c_+ \mathbf{v_+} [1 - \exp{(i \lambda_+ t)}] + c_- \mathbf{v_-} \left[1 - \exp{(i \lambda_- t)}\right].
\end{equation}
The RMS value of the eccentricity is then given by
\begin{equation}
\langle e_j^2 \rangle^{1/2} = \sqrt{\mathcal{E}_{fi}^2 + c_+^2 (v_+)_{j}^2 + c_-^2 (v_-)_{j}^2 },
\end{equation}
where $(v_{\pm})_j$ is the $j$-th component of vector $\mathbf{v}_{\pm}$ (Eq. \ref{eq:v_pm}). The equation above can be simplified to give the explicit expressions
for the RMS eccentricities of the two inner planets:
\begin{align}
\langle e_1^2 \rangle^{1/2} &= 
\sqrt{2} \left[\mathcal{E}_{f1}^2 + \frac{(-L_1 \mathcal{E}_{f1}^2 + L_2 \mathcal{E}_{f2}^2)\nu_{12}^2  - L_1 \mathcal{E}_{f1} \mathcal{E}_{f2}\Delta \omega \nu_{12}}{L_2 \Delta \omega^2 + 4L_1 \nu_{12}^2}  \right]^{1/2},
\label{eqn:gen_e1}
\end{align}
\begin{align}
\langle e_2^2 \rangle^{1/2} &= 
\sqrt{2} \left[\mathcal{E}_{f2}^2 + \frac{ (L_2 \mathcal{E}_{f1}^2 - L_1 \mathcal{E}_{f2}^2)\nu_{21}^2  - L_2 \mathcal{E}_{f1} \mathcal{E}_{f2}\Delta \omega \nu_{21}}{L_2 \Delta \omega^2 + 4L_1 \nu_{21}^2} \right]^{1/2}.
\label{eqn:gen_e2}
\end{align}

A comparison between the above expressions and the results of numerical integrations based on secular equations is shown in Fig. \ref{fig:fig2},
where we plot the RMS values of the planet eccentricities of a fiducial 2-planet inner system under the influence of a co-planar, eccentric giant perturber. As in the case of mutual inclinations, the eccentricities of the inner system fall into the three regimes characterized by strong inner planet coupling, resonance and weak inner planet coupling. We we elaborate on these in the following subsections.

\subsubsection{Resonance}
\label{sec:2.3.1}

When $\omega_{1} \simeq \omega_{2}$ (note that this is approximately equivalent to the condition $\epsilon_{12} \sim 1$), a potential resonance feature arises where large eccentricities can be excited in the inner planets, even for small $e_p$. If we take $\omega_{1} = \omega_{2}$, equations (\ref{eqn:gen_e1}) and  (\ref{eqn:gen_e2}) becomes
\begin{align}
\langle e^2_1 \rangle^{1/2}_{\mathrm{res}} &= 
\sqrt{2} \left[\frac{3L_1\mathcal{E}_{f1}^2 + L_2 \mathcal{E}_{f2}^2}{4L_1}  \right]^{1/2},
\label{eqn:res_e1}
\end{align}
\begin{align}
\langle e^2_2 \rangle^{1/2}_{\mathrm{res}} &= 
\sqrt{2} \left[\frac{5L_2\mathcal{E}_{f2}^2 - L_1 \mathcal{E}_{f2}^2}{4L_2}  \right]^{1/2}.
\label{eqn:res_e2}
\end{align}
We see that the eccentricity of planet 1 is boosted while the eccentricity of planet 2 is dampened near the resonance. The resonance feature is most pronounced when $L_2 \gg L_1$, To illustrate this, let $L_1 = 0$, then the forced eccentricity on the inner planet becomes
\begin{equation}
\mathcal{E}_{f1} = \left(\frac{\nu_{1p}}{\omega_1} + \frac{\nu_{2p}}{\omega_{2}} \right) \mathcal{E}_p, ~\quad (L_1 \ll L_2).
\end{equation}
If the inner planets both have zero initial eccentricity, then their eccentricity evolution is given by:
\begin{multline}
\mathcal{E}_1(t) =  \left[\frac{\nu_{1p}}{\omega_1} + \frac{\nu_{12}\nu_{2p}}{(\omega_1 - \omega_2)\omega_1}\right] \left[1 -\exp{(i \omega_1 t)}\right] \mathcal{E}_p \\ 
+ \left[\frac{\nu_{12}\nu_{2p}}{(\omega_1 - \omega_2)\omega_2}\right] \left[1 - \exp{(i\omega_2 t)}\right] \mathcal{E}_p,
\end{multline}
\begin{equation}
\mathcal{E}_2(t) = \left(\frac{\nu_{2p}}{\omega_2}\right)\left[1 - \exp{(i\omega_2 t)}\right] \mathcal{E}_p, ~\quad (L_1 \ll L_2).
\label{eqn:e_t_res}
\end{equation}
In this limiting case, according to the linear theory, at $\epsilon_{12} \simeq 1$ the eccentricity of the inner planet can become arbitrarily large, even for small initial values of $e_p$. In reality, the linear theory breaks down as $e_1$ becomes too large, and higher order terms will keep $e_1$ to a modest value.

An illustration of the resonance behavior can be seen in Fig. \ref{fig:fig2}. Systems with larger ratios of $m_2/m_1$ tend to exhibit pronounced resonance features, whereas for systems with more comparable masses, the feature is notably reduced.

\subsubsection{Strongly and Weakly Coupled Regime}

In the case where the mutual precession rates of the inner planets dominates over the influence of the perturber (i.e. $\omega_{12} \gg \omega_{1p} $ and $\omega_{21} \gg \omega_{2p})$, the general expressions (\ref{eqn:gen_e1}) and (\ref{eqn:gen_e2}) can be significantly simplified; we refer to this as the strongly coupled regime. In this regime, the two planets attain very similar forced eccentricities ($\mathcal{E}_{f1} \simeq \mathcal{E}_{f2}$), and as a result Eqs. (\ref{eqn:gen_e1}) and (\ref{eqn:gen_e2}) are dominated by their first terms. Explicitly, in this regime the RMS eccentricities are approximately given by
\begin{align}
\langle e_1^2 \rangle^{1/2} &\simeq \sqrt{2} \mathcal{E}_{f1} \simeq \sqrt{2} F_{11} e_p \nonumber \\
&\simeq \frac{5\sqrt{2} e_p}{4} \left[\frac{ \alpha_{12} + \beta_{12}  \left(\frac{m_2}{m_1}\right) \alpha_{12}^{-2} }{1 - \beta_{12}^2}\right] \left[\frac{3 \alpha_{12}}{b_{3/2}^{(1)}(\alpha_{12})}\right] \left(\frac{m_p}{m_2}\right) \left(\frac{a_2}{a_p}\right)^4  \label{eqn:strong_coupled_e1} \\
\langle e_2^2 \rangle^{1/2} &\simeq \sqrt{2} \mathcal{E}_{f2} \simeq \sqrt{2} F_{12} e_p \nonumber \\
&\simeq \frac{5\sqrt{2} e_p }{4} \left[\frac{ \beta_{12} \alpha_{12} +\left(\frac{m_2}{m_1}\right) \alpha_{12}^{-2}}{1 - \beta_{12}^2}\right] \left[\frac{3 \alpha_{12}}{b_{3/2}^{(1)}(\alpha_{12})}\right] \left(\frac{m_p}{m_2}\right) \left(\frac{a_2}{a_p}\right)^4,
\label{eqn:strong_coupled_e2}
\end{align}
where $\alpha_{12} = (a_1/a_2)$ and $\beta_{12} = \beta(a_1/a_2)$ (see Eq. \ref{eqn:beta12}).
In the other limiting case (i.e. weak coupling), when the precession rates of the inner planets driven by the perturber dominate the their mutual precession rates (i.e. $\nu_{1p} \gg \omega_{12}$ and $\nu_{2p} \gg \omega_{21}$), the terms of order $\nu_{12}/\omega_{1p}$ and $\nu_{21}/\omega_{2p}$ in Eqs. (\ref{eqn:gen_e1}) - (\ref{eqn:gen_e2}) can be dropped, and the final inner planets' RMS eccentricities are again given by $\sqrt{2} \mathcal{E}_f$. In this case, the planets precess independently of one another, and their eccentricities are given by Eqs. (\ref{eqn:e1_t}) and (\ref{eqn:e1_RMS}), i.e.
\begin{align}
\langle e_1^2 \rangle^{1/2} \simeq \sqrt{2} \left(\frac{5}{4}\right) \left(\frac{a_1}{a_p}\right) e_p, 
\label{eq:e1_weak} \\
\langle e_2^2 \rangle^{1/2} \simeq \sqrt{2} \left(\frac{5}{4}\right) \left(\frac{a_2}{a_p}\right) e_p.
\label{eq:e2_weak} 
\end{align}

Note that the criterion for the eccentricities of the inner planets to be strongly coupled or weakly coupled is related to the parameter $\epsilon_{12}$ (see Eq. \ref{eq:ep12}). For distant perturbers (i.e. $a_2 \ll a_p$), we generally have that $\nu_{1p} \sim \nu_{2p} \lesssim (\omega_{2p} - \omega_{1p})$. Therefore, the condition for strong eccentricity coupling is approximately $\max{(\nu_{1p},\nu_{2p})} \sim (\omega_{2p} - \omega_{1p}) \lesssim (\omega_{12} + \omega_{21})$, which is the same as $\epsilon_{12} \ll 1$.

Similarly, $\epsilon_{12} \gg 1$ implies that $\omega_{2p} - \omega_{1p} \gg \omega_{12} + \omega_{21}$. Since $\omega_{1p}, \omega_{2p} > (\omega_{2p} - \omega_{1p})$ while $\omega_{12}~, \omega_{21} < (\omega_{12} + \omega_{21})$, we find that $\epsilon_{12} \gg 1$ corresponds to the weak eccentricity coupling condition that $\omega_{2p} \gg \omega_{21}$ and $\omega_{1p} \gg \omega_{12}$. Although the correspondence between $\epsilon_{12}$ and the strong/weak coupling regimes is not exact, it serves as a useful dimensionless parameter for describing the dynamical evolution of the inner planet eccentricities.

The colored dashed curves in Fig. \ref{fig:fig2} compare the strong coupling and weak coupling approximations with both the full linearized theory (solid curves) as well as numerical secular integrations (dots). One can see that as $\epsilon_{12}$ approaches either very small or very large values, the limiting expression $\langle e_j^2 \rangle^{1/2} \simeq \sqrt{2} \mathcal{E}_{fj}$ becomes an increasingly more robust approximation for the full secular dynamics of the inner system.

Qualitatively, for $\epsilon_{12} \ll 1$, the inner planets are tightly coupled and their eccentricity vectors precess in tandem, and the eccentricity excitations are greatly muted; in this regime, $e_i \propto e_p m_p/a_p^4$. In the weak coupling regime ($\epsilon_{12} \gg 1$), $e_i \sim e_p(a_i/a_p)$. Note that in either limit, the scaling of $e_j$ is suppressed by factor $a_p^{-1}$ compared to the scaling for $\theta_{12}$; this is due to eccentricity oscillations being driven by the octupole (as opposed to the quadrupole) potential of the perturber.

However, there is one important difference between the eccentricity and inclination excitations for highly compact inner systems. As the inner planets become increasingly compact ($a_1/a_2 \rightarrow 1$), $\epsilon_{12} \rightarrow 0$ and the inner systems become essentially rigid and the mutual inclination induced by any perturber approaches zero. On the other hand, for highly compact planets even as $(a_2/a_1) \rightarrow 1$ the induced eccentricity approaches a finite value that scales with $(m_p/a_p^4)$. In other words, extremely compact systems that are strongly protected from mutual inclination excitations can still be somewhat susceptible to excitations in eccentricity.

This effect is shown in Fig. \ref{fig:fig3}, where the inner planet mutual inclinations and eccentricities are plotted for different values of $a_2/a_1$. We find that ceteris paribus (i.e. with $a_1, a_p$ fixed), as $(a_2/a_1) \rightarrow 1$, even a small decrease in $(a_2/a_1)$ leads to significant decreases in the inner planet mutual inclination excitations. For instance, as $(a_2/a_1)$ decreases from 1.3 to 1.2, $\theta_{12}$ decreases by a factor of $\sim 3$, whereas the changes in $e_1$ and $e_2$ are only $\sim 8 \%$. 


\section{Extension to Moderately Inclined and Eccentric Perturbers}
\label{sec:sec3}
\begin{figure*}
\includegraphics[width=0.95\linewidth]{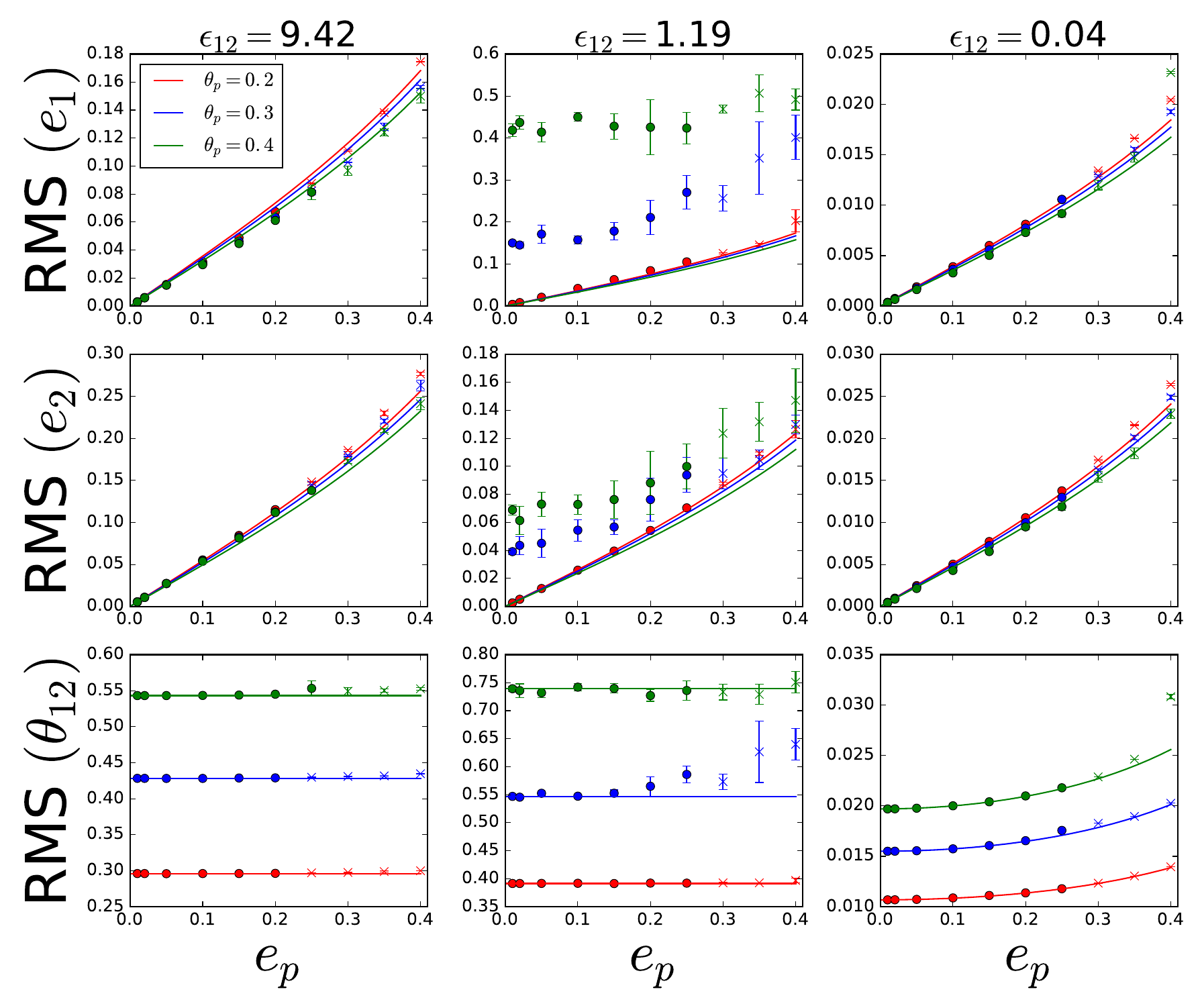}
\caption{RMS values of the inner planet eccentricities $e_1, e_2$ and their mutual inclination $\theta_{12}$ as a function of $e_p$ for a ``2-planet plus perturber system'' with $a_1 = 0.3$ au, $a_2 = 0.5$ au, $m_1 = 3M_{\oplus}$ and $m_2 = 5m_1$, perturbed by a $m_p = 5M_{J}$ planet. The panels, from left to right, represent three different perturber strengths $\epsilon_{12}$, which is varied by adjusting $a_p$. The different colors are for different values of $\theta_p$, with red, blue and green being $\theta_p = 0.2, 0.3, 0.4$ respectively. The points are the results of numerical secular equation integrations using Eqs. (\ref{eq:A1}) - (\ref{eq:A4}) while the solid lines are analytical results based on non-linear extensions to linear secular theory (Sec. \ref{sec:sec2}). For the left and center panels where $\epsilon_{12} > 1$, we obtain the solid curves from Eqs. (\ref{eq:nonlin_extend_weak_a}) - (\ref{eq:nonlin_extend_weak_c}), while for the right panel they were obtained from Eqs. (\ref{eq:nonlin_extend_strong_a}) - (\ref{eq:nonlin_extend_strong_c}). A point is marked with an `o' if the inner system is stable, and `$\times$' if it is unstable with respect to the stability criterion (Eq. \ref{eq:petro_stab}). Each point represents 3 different numerical secular simulations with otherwise identical initial parameters, except with the initial longitude of ascending node and longitude of perihelion sampled randomly in the interval $[0, 2\pi]$. }
\label{fig:fig4}
\end{figure*}

\begin{figure*}
\includegraphics[width=0.95\linewidth]{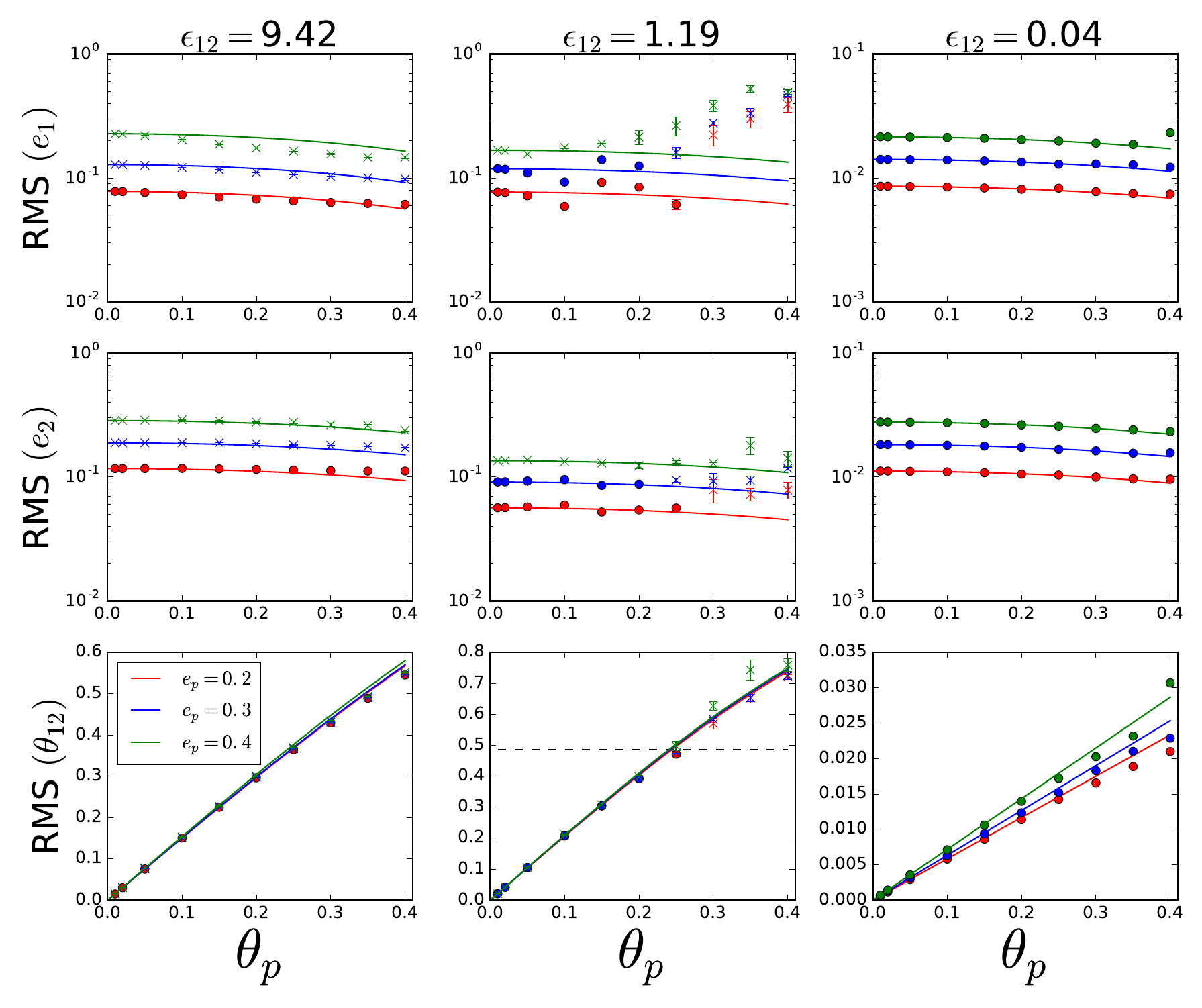}
\caption{Same as Fig. \ref{fig:fig4}, except that the results are plotted as a function of $\theta_p$. The different colors are for different values of $e_p$ with red, blue and green being $e_p = 0.2, 0.3, 0.4$ respectively. The dashed line in the center bottom panel corresponds to $\theta_{12} = 0.68$ rad. ($39^{\circ}$), the Lidov-Kozai critical angle. Note that when $\theta_{12} \le 0.68$ rad., the inner planets experience mutual Lidov-Kozai oscillations leading to large excitation of $e_1$ (see the middle column).}
\label{fig:fig5}
\end{figure*}

When both $e_p$ and $\theta_p$ are significant, the linearized secular
theory breaks down and one must resort to secular or N-body numerical
integrations. However, note that for inner planets in the strong coupling
regime, the planets maintain small eccentricities and mutual
inclinations. As a result, for this region of parameter space the
evolution is non-linear only in the variables $e_p$ and
$\theta_p$. This allows us to extend the regime of validity of the
result based on linearized secular theory by substituting
the linear scalings of $e_p, \theta_p$ with the appropriate non-linear scalings.

\begin{figure*}
\includegraphics[width=0.95\linewidth]{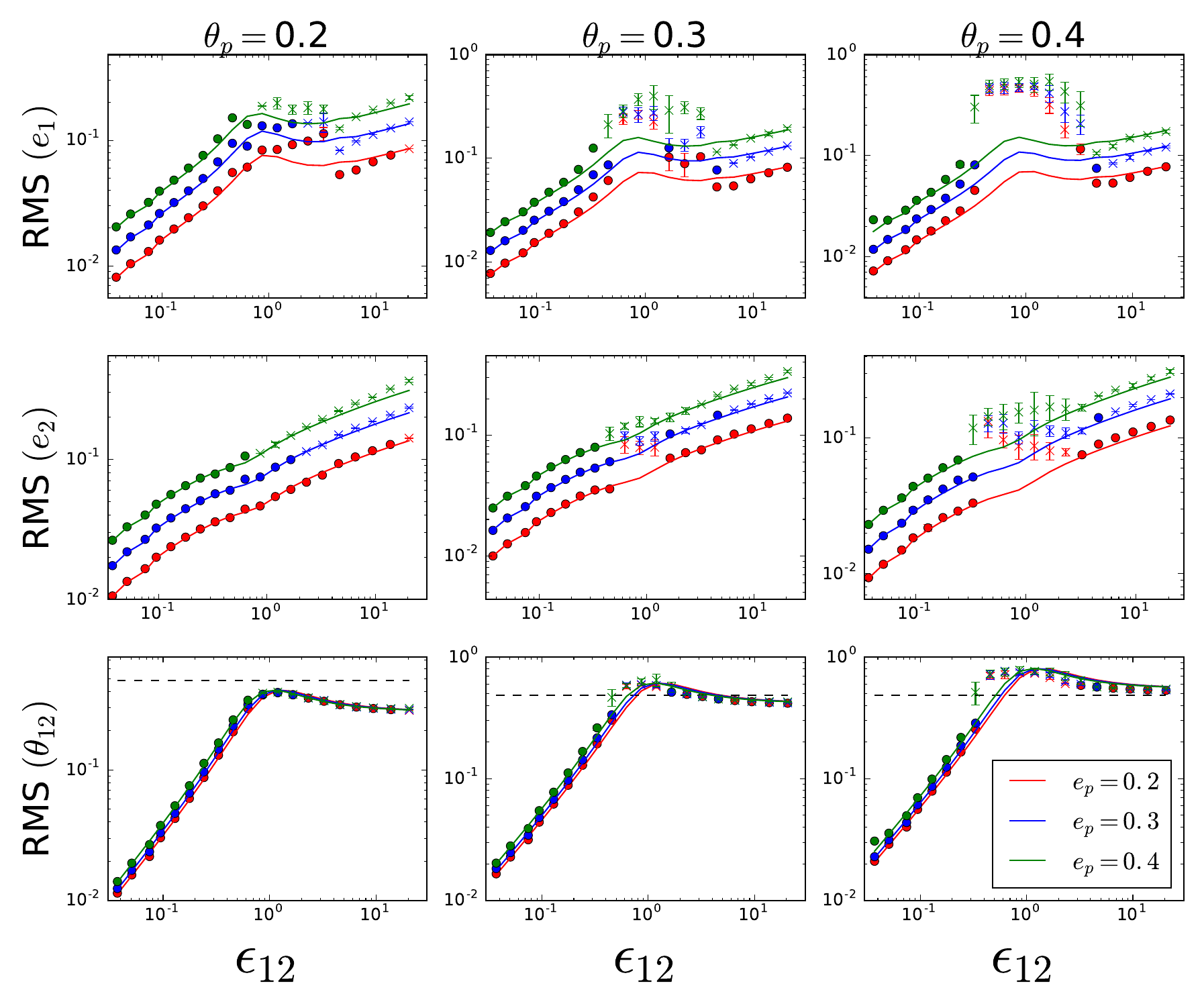}
\caption{Same as Fig. \ref{fig:fig4}, except that the results are plotted as a function of $\epsilon_{12}$. The variation of $\epsilon_{12}$ is achieved by fixing $m_p = 5M_J$ while varying $a_p$. The different colors are for different values of $e_p$, with red, blue and green being $e_p = 0.2, 0.3, 0.4$ respectively. The panels from left to right have different values of $\theta_p = 0.2, 0.3, 0.4$ respectively. For the bottom three panels, the solid lines are derived from Eq. (\ref{eq:theta12}), but with $\omega_{jp}$ replaced by $\tilde{\omega}_{jp}$ (Eq. \ref{eq:omega_tilde}). The horizontal dashed line corresponds to $\theta_{12} = 0.68$ rad. }
\label{fig:fig6}
\end{figure*}

\subsection{Secular Equations}
\label{sec:secular_equations}

The secular evolution multi-planet systems can be studied using
different approaches, each having its own regime of validity.
The standard Laplace-Lagrange theory \citep[see][the vector form is used in Section \ref{sec:sec2}]{MurrayDermott} assumes small eccentricities and
inclinations for all planets (and perturbers), but allows arbitrary
ratio of semi-major axes between planets, as long as the
system is dynamically stable. On the other hand, for hierarchical
systems, one can expand the interaction potential between planets (and
perturbers) in terms of the (small) ratio of semi-major axes, while
allowing for arbitrary eccentricities and inclinations -- this multipole
expansion approach has a long history \citep[e.g.][]{Lidov1962, Kozai1962, Ford2000, Laskar2010}.
We derive our results from the secular equations based on full,
doubly-averaged multipole expansion of the disturbing potential up to
the octupole order. We use the vector form of the secular equations 
of motion \citetext{\citealt{Liu2015}; see also \citealt{Boue2014}, \citealt{Petrovich2015_Secular}},
expressed in terms of the dimensionless angular momentum vector ${\jvec}_j$ 
and eccentricity vector $\evec_{j}$ of each planet (see Eq. \ref{eqn:evec_jvec}). While these equations of motion accurately account for the interaction between a planet in the inner system and the distant perturber, they
become a poor approximation when describing the interaction between the inner,
closely spaced planets. We therefore use a modified version of the
equations of motion that hybridizes the Lagrange-Laplace theory with
the multipole expansion results by introducing appropriate Laplace
coefficients in place of the usual $a_<^n/a_>^{n+1}$ terms in the
multipole expansion (see Appendix \ref{sec:appendix}).

While our hybrid equations of motion are formally nonlinear in terms of the
eccentricities and inclinations of all planets, they are valid only when
the inner planets have small eccentricies and mutual inclinations.
For systems where the inner planets are strongly packed and $e_p,
~\theta_p$ of the external perturber are modest (e.g. $\lesssim 0.4$),
the inner planet eccentricities and mutual inclinations will be small,
and the equations are non-linear only in $e_p$ and $\theta_p$.  (In
the case when the inner planets develop large eccentricities and/or
mutual inclinations, dynamical instability is likely to set in and the
secular theory breaks down.) As a result, we can treat the inner
planets as behaving linearly, but subject to external forcing with a
non-linear dependence on $e_p, ~\theta_p$.

To the leading order, the evolution of $\jvec_j,~ \evec_j$ of an inner
planet due to the perturber is given by (see Eqs. \ref{eq:A1} - \ref{eq:A2})
\begin{equation}
\left(\frac{d{\jvec_{j}}}{dt}\right)_p \simeq \frac{\omega_{jp}\cos{\theta_{jp}}}{(1-e_p^2)^{3/2}}  (\jvec_j\times\nvec_p),
\label{eq:j1vec_approx} 
\end{equation}
\begin{align}
\left(\frac{d{\evec_{j}}}{dt}\right)_p &\simeq \frac{\omega_{jp}}{(1-e_p^2)^{3/2}}  \left[\cos{\theta_{jp}}(\evec_j\times\nvec_p) - 2(\evec_j \times \jvec_j) \right] \nonumber \\
& \qquad \qquad -\frac{5\nu_{jp}}{4(1-e_p^2)^{5/2}}\left[\cos^2{\theta_{jp}}-\frac{1}{5}\right](\jvec_j \times \evec_p) \nonumber \\
&\simeq -\frac{\omega_{jp}\cos{\theta_{jp}}}{(1-e_p^2)^{3/2}} (\evec_j\times\nvec_p) \nonumber \\
& \qquad \qquad -\frac{5\nu_{jp}}{4(1-e_p^2)^{5/2}}\left[\cos^2{\theta_{jp}}-\frac{1}{5}\right](\jvec_j \times \evec_p),
\label{eq:e1vec_approx}
\end{align}
where in the last equation we have replaced the term $(\evec_j \times \jvec_j)$  with its time-average by taking $\jvec_j \approx \langle \jvec_j(t) \rangle = \cos{\theta_{jp}} \nvec_p$. The terms proportional to $(\evec_j \times \nvec_p)(\jvec_j \times \nvec_p)$ were dropped as they are proportional to $e_j$ which is assumed to be small. We now define:
\begin{align}
  \tilde{\omega}_{jp} &\equiv \omega_{jp} \left[\frac{\cos{\theta_{jp}}}{(1-e_p^2)^{3/2}}\right],   \label{eq:omega_tilde}  \\
  \tilde{\nu}_{jp} &\equiv \nu_{jp}
  \left[\frac{5\cos^2{\theta_{jp}} - 1 }{4(1-e_p^2)^{5/2}}\right].
  \label{eq:nu_tilde}
\end{align}
We then obtain a modified version of the Laplace-Lagrange evolution equations for the eccentricity vector $\evec_j$ and
unit angular momentum vector $\jvec_j$ (with $k = 1,2,3...N$, not including the perturber):
\begin{align}
\frac{d \evec_j}{dt} &= - \sum_{k \neq j} \omega_{jk} (\evec_j \times \jvec_k) - \sum_{k \neq j} \nu_{jk} (\jvec_j \times \evec_k) 
- \tilde{\omega}_{jp} (\evec_j \times \jvec_p) - \tilde{\nu}_{jp} (\jvec_j \times \evec_p),
\label{eq:LL_e_nonlin}\\
\frac{d \jvec_j}{dt} &=  \sum_{k \neq j} \omega_{jk} (\jvec_j \times \jvec_k)
+ \tilde{\omega}_{jp} (\jvec_j \times \jvec_p).
\label{eq:LL_t_nonlin}
\end{align}
The above equations are analogous to the linearized Laplace-Lagrange theory (Eq. \ref{eq:LL_e} and \ref{eq:LL_l}), except with modified quadrupole and octupole precession frequencies $\tilde{\omega}_{jp}$ and $\tilde{\nu}_{jp}$.

\subsection{Results for ``Two Planets + Perturber'' System}

We can now extend the analytical results of Section \ref{sec:sec2} to finite $e_p$ and $\theta_p$. Specifically, the mutual inclination $\theta_{12}$ of the ``two planets + perturber'' system can be obtained using Eq. (\ref{eq:theta12}), except with $\omega_{1p}$ and $\omega_{2p}$ replaced by $\tilde{\omega}_{1p}$ and $\tilde{\omega}_{2p}$, respectively. Similarly, the inner planet eccentricities can be computed using Eqs. (\ref{eqn:gen_e1}) - (\ref{eqn:gen_e2}), except with $\omega_{jp}$ and $\nu_{jp}$ replaced by $\tilde{\omega}_{jp}$ and $\tilde{\nu}_{jp}$ for $j \in \{1, 2\}$.

When the inner planets are strongly coupled $(\epsilon_{12} \ll 1)$, the extension to mildly eccentric and misaligned perturbers can be simplified even further. Substituting the frequencies from Eqs. (\ref{eq:omega_tilde}) - (\ref{eq:nu_tilde}) into Eqs. (\ref{eq:theta12}), (\ref{eqn:gen_e1}) and (\ref{eqn:gen_e2}), and making an expansion to second order in $\theta_p$ we have
\begin{align}
\langle         e_1^2 \rangle^{1/2}(e_p, \theta_p) &\simeq   \left[\frac{1 - 5\theta_p^2/4}{(1 -e_p^2)^{5/2}}\right] \langle e_1^2 \rangle^{1/2}_{\mathrm{lin}} ,
\label{eq:nonlin_extend_strong_a} \\
\langle         e_2^2 \rangle^{1/2}(e_p, \theta_p) &\simeq   \left[\frac{1 - 5\theta_p^2/4}{(1 -e_p^2)^{5/2}}\right] \langle e_2^2 \rangle^{1/2}_{\mathrm{lin}}, \label{eq:nonlin_extend_strong_b} \\
\langle \theta_{12}^2 \rangle^{1/2}(e_p, \theta_p) &\simeq   \left[\frac{1 - \theta_p^2/2}{(1 -e_p^2)^{3/2}}\right]  \langle \theta_{12}^2 \rangle^{1/2}_{\mathrm{lin}},
\label{eq:nonlin_extend_strong_c}
\end{align}
where the ``linear'' expressions for $\langle e_1^2 \rangle^{1/2}_{\mathrm{lin}}$, $\langle e_2^2 \rangle^{1/2}_{\mathrm{lin}}$ and $ \langle \theta_{12}^2 \rangle^{1/2}_{\mathrm{lin}}$ are given by Eqs. (\ref{eqn:e1_RMS}), (\ref{eqn:gen_e2}) and (\ref{eq:theta12}), respectively. 

When the inner planets are weakly coupled $(\epsilon_{12} \gg 1)$, a similar simplification can be made. However, we caution that the underlying assumptions that $e_j$ and $\theta_{jk}$ are small may no longer be valid in this regime, and the non-linear extensions in this case may not be fully justified. Nonetheless, we include them below for completeness:
\begin{align}
\langle         e_1^2 \rangle^{1/2}(e_p, \theta_p) &\simeq  \left(\frac{1 - 3\theta_p^2/4}{1 -e_p^2}\right) \langle e_1^2 \rangle^{1/2}_{\mathrm{lin}},
\label{eq:nonlin_extend_weak_a} \\
\langle         e_2^2 \rangle^{1/2}(e_p, \theta_p) &\simeq  \left(\frac{1 - 3\theta_p^2/4}{1 -e_p^2}\right) \langle e_2^2 \rangle^{1/2}_{\mathrm{lin}}, \label{eq:nonlin_extend_weak_b} \\
\langle \theta_{12}^2 \rangle^{1/2}(e_p, \theta_p) &\simeq  \langle \theta_{12}^2 \rangle^{1/2}_{\mathrm{lin}},
\label{eq:nonlin_extend_weak_c}
\end{align}
where the ``linear'' expressions for $\langle e_j^2 \rangle^{1/2}_{\mathrm{lin}}$ are given by Eq. (\ref{eqn:e1_RMS}), and $ \langle \theta_{12}^2 \rangle^{1/2}_{\mathrm{lin}}$ is given by Eq. (\ref{eq:theta_12_weak}).

Our results are summarized in Figs. \ref{fig:fig4} - \ref{fig:fig6}, in which we show the RMS values of the inner planet eccentricities $e_1$ and $e_2$ and mutual inclination $\theta_{12}$ as a function of $e_p$, $\theta_p$ and $a_p$ (varied by varying $\epsilon_{12}$) respectively. In each figure, we compare the results of our non-linear extension to linear theory (solid curves) with numerical secular integrations (points). The linear theory extension to mild $\theta_p$ and $e_p$ appears to agree well with numerical secular integrations up to values of $e_p$ and $\theta_p \sim 0.4$. One case where our analytical expressions agree poorly with secular integrations occur when $m_1 \ll m_2$, $\epsilon_{12} \simeq 1$ and $\theta_{12} \ge 39^{\circ}$, seen most prominently for the green points in the middle panels of Figs. \ref{fig:fig4} and \ref{fig:fig5} and the top panels of Fig. \ref{fig:fig6}. We elaborate on this feature in the next subsection.

\subsection{Resonance Feature and Internal Lidov-Kozai Oscillations}
\label{sec:kozai_lidov_like}

In cases with mild $e_p, \theta_p$, for certain parameters the previously noted resonance at $\epsilon_{12} \simeq 1$ takes on a richer behavior (see Section \ref{sec:2.3.1}). We find that for systems with $\epsilon_{12} \simeq 1$, whenever $\theta_p$ is sufficiently large such that the inner planets attain the critical mutual inclination angle for Lidov-Kozai oscillations ($\theta_{12} \ge 0.68 ~\mathrm{rad}. = 39^{\circ}$) \citep[see][]{Lidov1962,Kozai1962}, the inner planets can experience dramatic growth in eccentricity regardless of the value of $e_p$, in a fashion analogous to secular oscillations first described by \cite{Lidov1962} and \cite{Kozai1962}. This is most clearly seen in the middle column of Fig. \ref{fig:fig5}. The rise in $\langle e_1^2 \rangle^{1/2}$ coincides with the inner planets obtaining a mutual inclination greater than $39^{\circ}$. This behavior is not predicted by the linearized secular theory, which always yields $e_1 \lesssim e_p$ for $e_p \ll 1$.

In order for the Lidov-Kozai-like oscillations between the inner planets to occur, we find the three following criteria are required: The innermost planet should be less massive than the outer one (i.e. $m_1 \lesssim m_2$); the ``2 planets plus perturber'' system should be near the secular resonance, with $\epsilon_{12} \simeq 1$; the misalignment angle $\theta_p$ should be smaller than the Kozai critical angle ($\theta_p \lesssim 0.68 ~\mathrm{rad.}$), but sufficiently large to induce the inner pair to have a mutual misalignment greater than the Lidov-Kozai critical angle ($\theta_{12} \simeq 0.68 ~\mathrm{rad}.$)\footnote{Note that for $\theta_p \gtrsim 0.68$ rad., the perturber will drive the inner planets into conventional Lidov-Kozai oscillations, resulting in close encounters and/or collisions between the inner planets as their orbits would eventually cross.}.

In our numerical investigations, we find that both our hybrid secular numerical algorithm and N-body calculations exhibit this behavior, although the N-body results deviate somewhat from our secular predictions. We find that such mutual Lidov-Kozai cycles generally require larger $\theta_p$ in the N-body simulations and take on milder behaviors (see Section \ref{sec:sec4}).

\section{Comparison with N-body Integrations}
\label{sec:sec4}

\begin{figure*}
\includegraphics[width=0.95\linewidth]{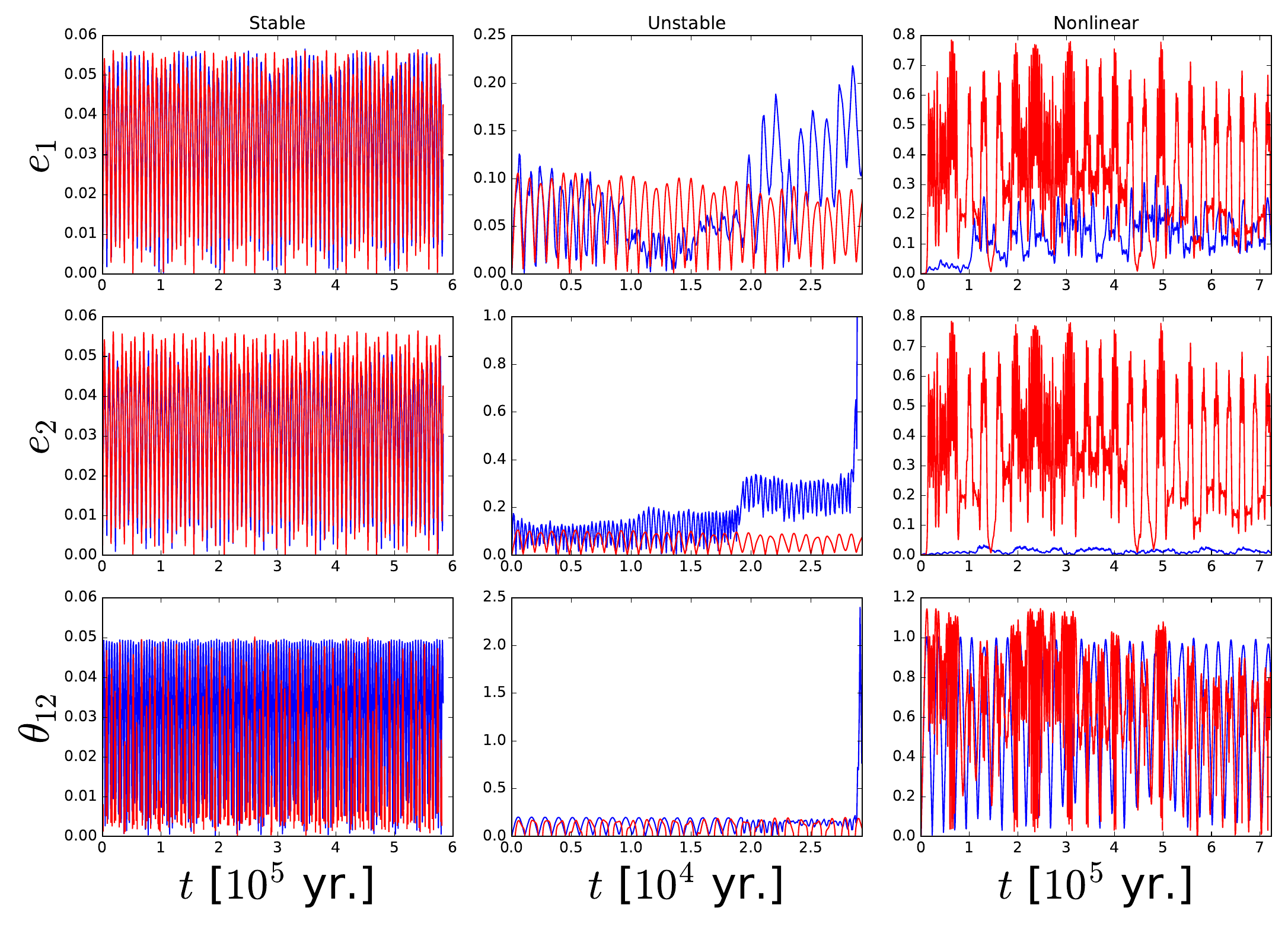}

\caption{Sample evolution of ``two-planet + perturber'' system using N-body integrations. From top to bottom, the y-axis shows $e_1$, $e_2$, and $\theta_{12}$. From left to right are three different scenarios corresponding to the stable, unstable and nonlinear (Lidov-Kozai-like oscillation) regimes. The red curve shows results based on secular integration while the blue curve are from N-body integration using the same initial parameters.  Left: an example where the secular hybrid algorithm matches closely with N-body integrations. For this particular model, $a_1, a_2, a_p$ are 0.3, 0.39 and 1.6 au respectively, while $m_1 = m_2 = 3 M_{\oplus}$ and $m_p = 3 M_J$. The perturber has $e_p = a_p = 0.1$. Center: Same as left, except here $a_p = 1.26$ au. This is an example of the inner two planets driven into dynamical instability, as a result of eccentricity excitation by the perturber. Right: An example of Kozai-like oscillations. Here, $a_1, a_2, a_p$ are $0.3, 0.45, 1.94$ au respectively, such that $\epsilon_{12} \simeq 1$. The planets have masses $m_1 = 0.6 M_{\oplus}$, $m_2 = 3M_{\oplus}$ and $m_p = 3M_J$, and the perturber has $m_p = 3M_J$, $e_p = 0.02$ and $\theta_p = 0.4$. }
\label{fig:fig7}
\end{figure*}

\begin{figure*}
\includegraphics[width=0.95\linewidth]{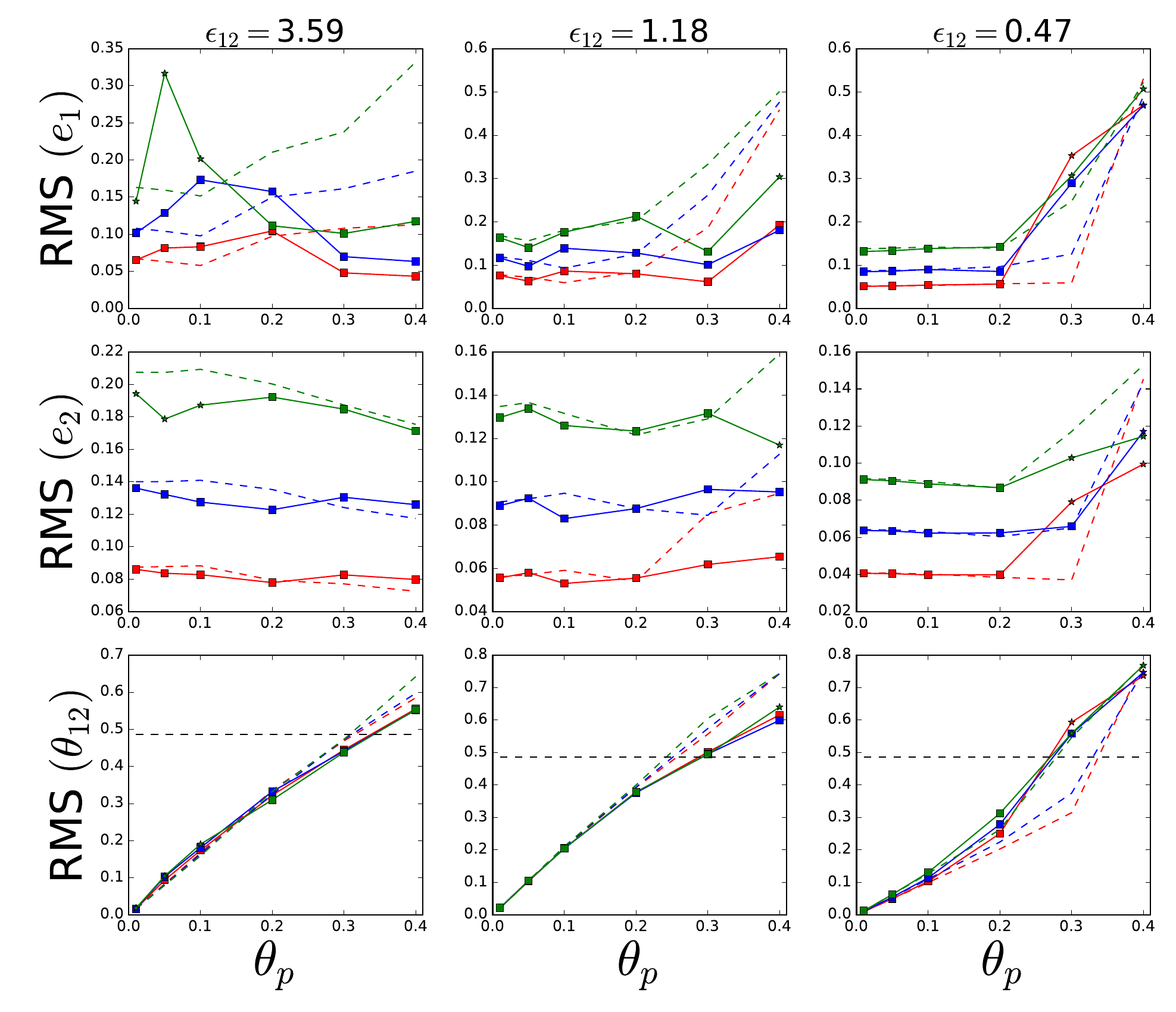}
\caption{RMS values of planet eccentricities and inclinations for a ``2-planet + perturber'' system with $a_1 = 0.3$ au, $a_2 = 0.5$ au, $m_1 = 0.6 M_{\oplus}$ and $m_2 = 3M_{\oplus}$, perturbed by a $m_p = 5M_J$ planet. The panels, from left to right, represent three different perturber strengths $\epsilon_{12}$, which is varied by adjusting $a_p$. The different colors are for different values of $e_p$, with red, blue and green corresponding to $e_p = 0.2, ~0.3$, and $0.4$ respectively. The solid curves are the results based on N-body integrations while the dashed curves are the results of hybrid secular equations. Systems that are stable during the integration were marked with a filled square for the solid curves, while those that were unstable with respect to orbit crossings are marked with a star. The dashed horizontal lines on the bottom panels corresponds to $\theta_{12} = 0.68$ rad. }
\label{fig:fig8}
\end{figure*}

\begin{figure*}
\includegraphics[width=0.95\linewidth]{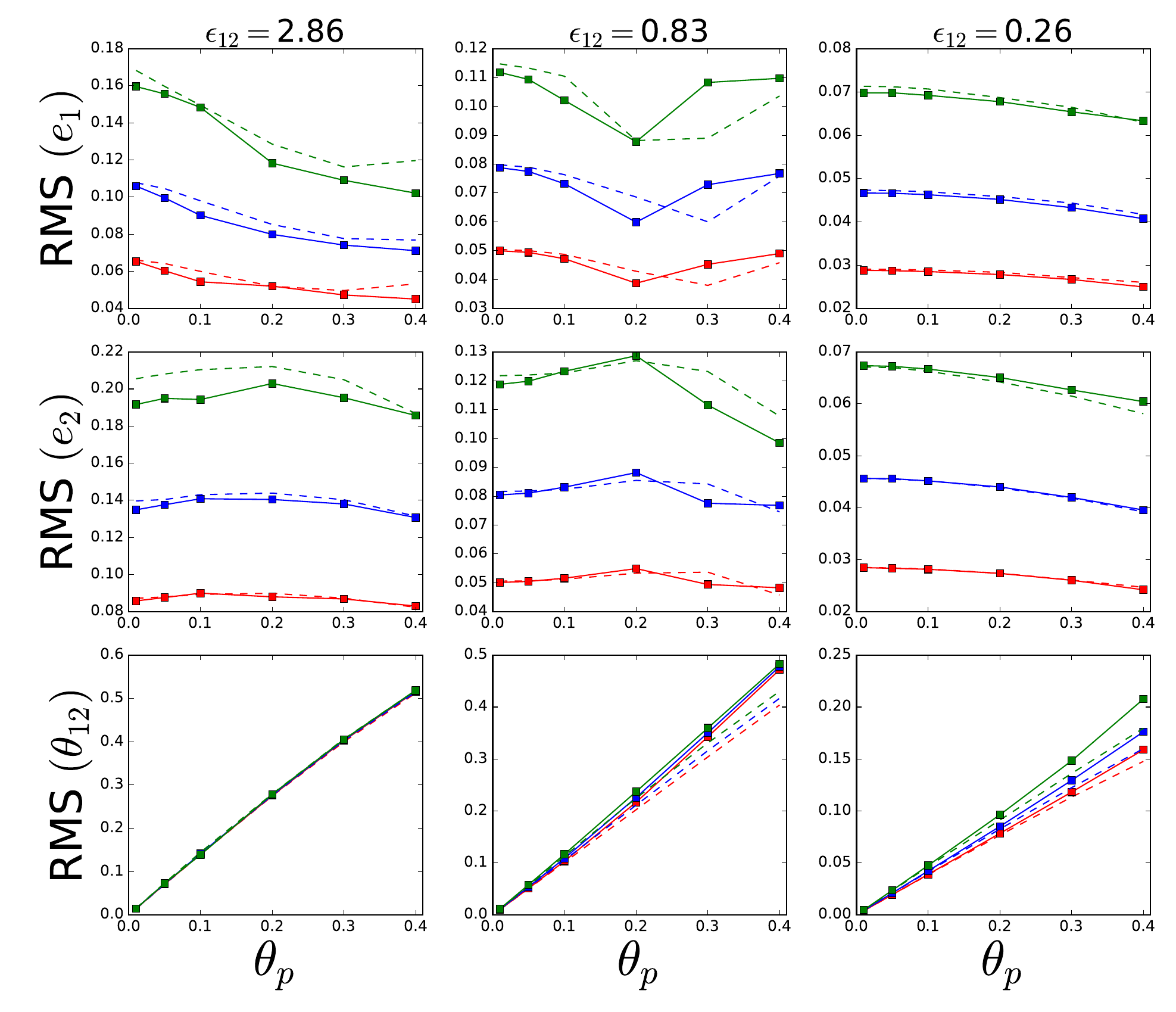}
\caption{Same as Fig. \ref{fig:fig8}, except that $m_1 = m_2 = 3 M_{\oplus}$.}
\label{fig:fig9}
\end{figure*}

We compare our results based on secular equations with N-body simulations by computing the same systems with REBOUND, using the WHFast integrator \citep{ReinLiu2012, ReinTamayo2015}. We chose $a_1 = 0.3$ au and $a_2 = 0.5$ au, with $a_p$ varying between $0.9 - 6$ au. We select our timestep such that $dt$ is equal to 1/40 of the orbital periods of the innermost planet, and we integrate our systems up to $10^6$ yr or until one of the planets is ejected. The planets were taken to be point masses and physical collision were ignored. The planet masses are $m_2 = 3 M_{\oplus}$, $m_p = 5M_{J}$ and $m_1$ was either $0.5$ or $3.0 M_{\oplus}$. The first case (with $m_1 = 0.5 M_{\oplus}$) represents a scenario where the secular resonance feature can significantly boost $e_1$ and $\theta_{12}$. We start the integration with the inner planets in circular, co-planar orbits while the perturber has inclinations and eccentricities taken from the set [0.01, 0.02, 0.05, 0.1, 0.2, 0.3, 0.4]. The RMS eccentricities and inclinations are calculated over the timespan from the start until the end of the simulation.

The results of our comparison between N-body integrations and hybrid secular equation integrations are shown in Figs. \ref{fig:fig7} - \ref{fig:fig9}. These results can be generally divided into three regimes, corresponding to the three columns of Fig. \ref{fig:fig7}:

$\bullet$ In the first regime (left panels of Fig. \ref{fig:fig7}), the inner planets undergo steady oscillations in their eccentricities and mutual inclinations; we call this the ``stable'' regime. The planet eccentricities and mutual inclinations ($e_1, ~e_2, ~\theta_{12}$) tend to remain small in this regime, and there is excellent agreement between the results of our hybrid secular equations and N-body integrations. Note that while the oscillation amplitudes of the planet eccentricities and inclinations agree between the two methods, there is also a notable difference in the phase of the oscillations; this is in agreement with other studies and tends to usually be the case for systems in the stable regime.

$\bullet$ In the second regime (middle column of Fig. \ref{fig:fig7}), the perturber drives the inner system into gravitational instability, leading to close encounters and/or orbit crossings between planets that are not captured by secular dynamics; we call this the `unstable' regime. This regime typically corresponds to systems with $\epsilon_{12} > 1$ and modest $e_p$. In this scenario, whereas the hybrid secular integrations show stable oscillations for the planet eccentricities and mutual inclinations, the N-body simulations feature sudden and drastic growth in the eccentricities and mutual inclinations of the planets, eventually leading to planet collisions or ejections. Such planet systems would appear to be stable in our integrations based on secular equations, but are in reality unstable in the long-term. 

Since secular methods are unsuitable for the study of these unstable systems, we caution that some of our secular results which leads to systems that undergo close encounters may produce misleading results. One way to filter out such potentially unstable systems is to use a stability criterion based on orbital parameters to identify systems that are gravitationally unstable. \cite{Petrovich2015} found empirically that the criterion for a pair of planets on somewhat co-planar orbits ($\theta_{12} \lesssim 39^{\circ}$) to be stable for all time is given by:
\begin{equation}
\frac{a_{1}(1 - e_1)}{a_2(1+e_2)} \le 1.2.
\label{eq:petro_stab}
\end{equation}
Note that the above criterion was based on ensembles of numerical N-body integrations with planet ratios $\mu \equiv m/M_{*}$ between $10^{-2} - 10^{-4}$, and therefore is an overestimate for the stability criterion for planets with masses more comparable to super-Earths ($\mu \sim 10^{-5}$); nevertheless, we adopt it a a conservative estimate. In this study we adopt the above stability criterion for 2-planet systems and check our secular integrations for instability against this criteria. We found that for some of our numerical secular equations integrations, an external perturber could indeed excite the inner planets into instability for large enough $\epsilon_{12}$ and $e_p$. The parameters leading to this instability is marked with an `$\times$' in Figs. \ref{fig:fig4} - \ref{fig:fig6}. We caution that secular theory cannot adequately describe the dynamics of these systems and one should resort to full N-body simulations.

$\bullet$ A third regime of final outcomes occurs when the inner planets undergo Lidov-Kozai-like oscillations (right column of Fig \ref{fig:fig7}), discussed in Section \ref{sec:kozai_lidov_like}. In this regime, the results of secular integrations tend to be qualitatively similar to the N-body integrations, but with qualitative differences in the oscillation amplitudes of the planet eccentricities $e_1$ and $e_2$. In comparison to the secular integrations, N-body integrations generally feature far milder eccentricity growth in the inner planets. In the example shown on the right of Fig. \ref{fig:fig7}, whereas the secular integrations predict $e_1$ and $e_2$ to reach values of $\sim 0.8$ and $\sim 0.14$ respectively, the N-body integrations showed much smaller values of $e_1 \sim 0.15$ and $e_2 \sim 0.02$ respectively. The overall trend is most clearly seen in the middle column of Fig. \ref{fig:fig8}, where even though both the secular integrations and N-body integrations show a steep increase in $\langle e_1^2 \rangle^{1/2}$ as $\theta_{12}$ increases beyond $39^{\circ}$, the magnitude of $\langle e_1^2 \rangle^{1/2}$ seen in N-body integrations is generally smaller than the secular integrations. The discrepancy is likely due to the presence of higher order corrections to the secular equations, which were not captured by our expansion up to octupole order. 

Figs. \ref{fig:fig8} and \ref{fig:fig9} show a comparison of the final RMS planet eccentricities and mutual inclinations obtained from our hybrid secular equations (dashes) versus N-body simulations (stars and filled squares); planet systems that became unstable in the N-body simulations due to the effect of the perturber were marked with a star, otherwise they were marked by filled squares. For Fig. \ref{fig:fig8}, the planet masses were chosen to allow for resonance features and Lidov-Kozai-like oscillations to occur, by setting $m_1 = 0.6 M_{\oplus}$ and $m_2 = 3 M_{\oplus}$. The columns represent different coupling regimes, with the left, center and right panels corresponding to weak, resonant and strong coupling respectively, while the different colors represent different perturber eccentricities (with red, blue and green being $e_p = 0.2, ~0.3, ~0.4$ respectively). When $\epsilon_{12}$, $e_p$ and $\theta_p$ are small, there is good agreement between the results of hybrid secular equations and N-body integrations as the inner planets remain in the stable regime. For the case of $\epsilon_{12} \gtrsim 1$, the inner planets can be driven into the unstable regime (see, e.g. the green curve on the left-side panels), and there are more substantial deviations between our hybrid secular equations and N-body integrations. When the inner planets achieve $\theta_{12} \le 0.68$ rad. (delineated by the dashed line on the bottom panels), Lidov-Kozai-like oscillations develop and the agreement between hybrid secular equations and N-body simulations become poor.

In Fig. \ref{fig:fig9}, we show the same comparisons as Fig. \ref{fig:fig8}, except with the inner planets having equal masses ($m_1 = m_2 = 3M_{\oplus}$) to prevent the development of resonance effects or Lidov-Kozai-like oscillations. In this case, we find strong agreements between the hybrid secular equations and N-body integrations across the range of parameters.

\begin{figure*}
\includegraphics[width=0.95\linewidth]{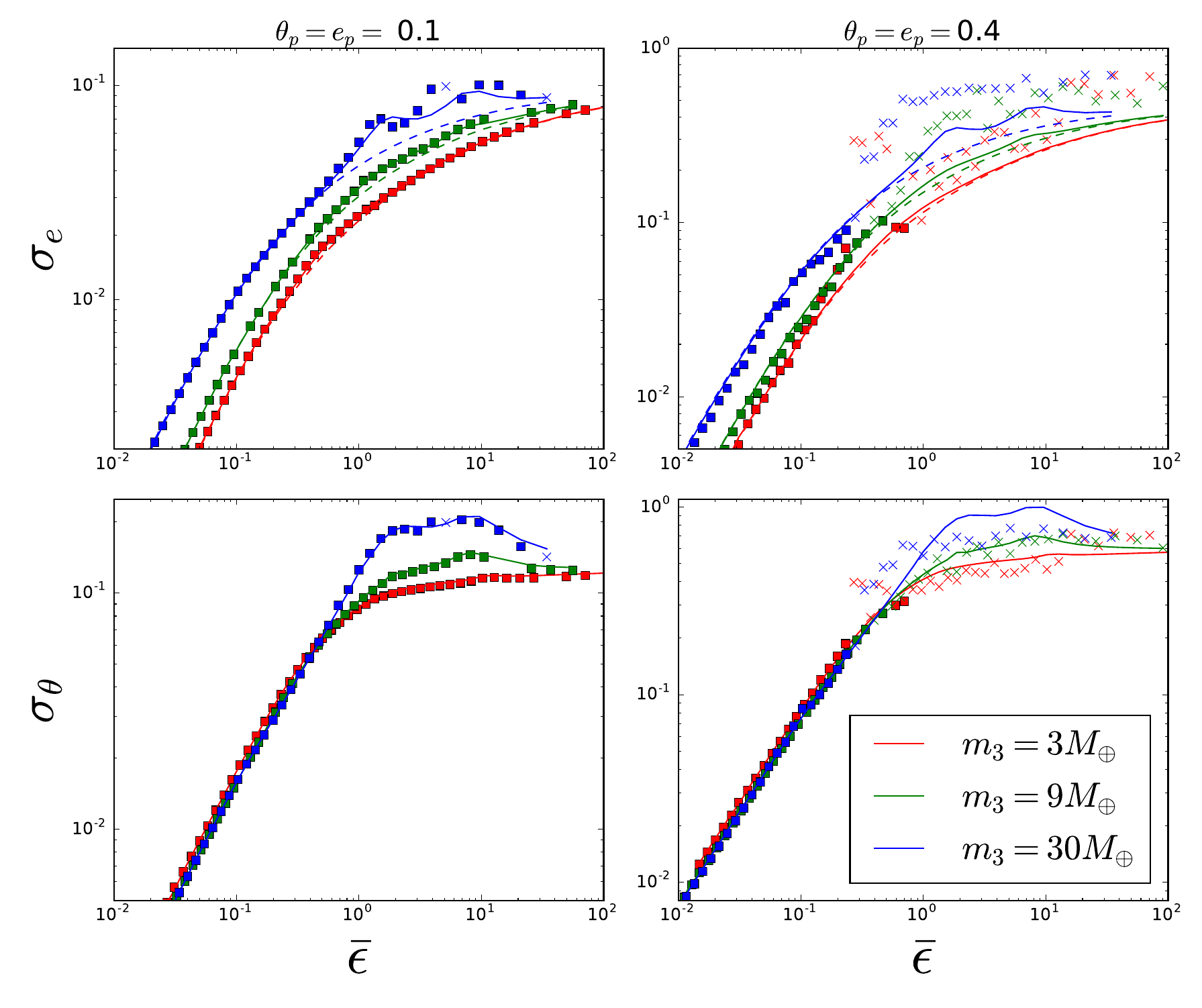}
\caption{Comparison of the RMS values of planet eccentricities and inclinations for a hypothetical 4-planet system under the influence of an inclined, eccentric perturber, computed from secular theory (solid and dashed curves) and N-body simulations (`$\times$' and squares). The planets have masses $m_1 = m_2 = m_4 = 3M_{\oplus}$, and the 3rd planet is the `dominant' one with $m_d = m_3$ equal to $3, ~9,$ and $30 M_{\oplus}$ for the red, green and blue curves respectively. The semi-major axes of the four inner planets are $[0.1, 0.15, 0.25, 0.4]$ au, while $\bar{\epsilon}$ is varied by varying $a_p$. Points that are marked with filled squares represent systems stable against orbit crossings, while systems that have undergone orbit crossings are marked with an `$\times$'. From the top panel, the y-axis is the RMS eccentricity averaged over all planets, as function of mean coupling parameter $\bar{\epsilon}$. The left panel represents a ``linear'' case with $e_p = \theta_p = 0.1$ while the right panel has $e_p = \theta_p = 0.4$. The colored dashed lines are given by the forced eccentricity only (Eq. \ref{eq:rms_e_strong_coupling}), while the solid curves are analytical estimates based on linear theory, given by Eq. (\ref{eq:rms_e_gen}) on the left panel and using the mild $e_p$ extension (Eq. \ref{eq:nonlin_extend_strong_multi_e}) on the right panel. The bottom panels are similar to the top panels, except the mean pairwise mutual inclination $\sigma_\theta$ is plotted on the y-axis. The solid curves are obtained from Eq. (\ref{eq:theta_p_multi}) for the left panel, with the nonlinear extension Eq. (\ref{eq:nonlin_extend_strong_multi_th}) being used for the right panel.}
\label{fig:fig10}
\end{figure*}

\section{N > 2 Inner Planets + Perturber}
\label{sec:sec5}

The results of the above sections, applicable for inner systems with $N = 2$ planets, can be generalized to systems with more than two inner planets. Consider a system of $N > 2$ inner planets of masses $m_i$ ($i =
1,2,3...,N$) and semi-major axes $a_i$ ($a_1 < a_2 < ... <
a_N$), accompanied by a giant planet (or stellar) perturber (with
$m_p \gg m_i$, semi-major axis $a_p \gg a_i$, inclination angle
$\theta_p$ and eccentricity $e_p$). The evolution equations for the eccentricity and inclination vectors of the $j$-th planet $\evec_j$ and $\jvec_j$ are given by Eqs. (\ref{eq:A1}) - (\ref{eq:A4}). Given an inner system, how do the inner planet eccentricities and mutual inclinations change as a function of $e_p$, $\theta_p$, $m_p$ and $a_p$?

For systems with many planets it is useful to consider the averaged dynamical quantities of all planets. We define $\sigma_{\theta}$ as the RMS time-averaged mis-alignment angle between all planet pairs:
\begin{equation}
\sigma_{\theta} \equiv \sin^{-1}{\left[\left( \frac{1}{N(N-1)}  \sum_j \sum_{k \neq j} \langle |\nvec_j \times \nvec_k|^2 \rangle \right)^{1/2}\right]}.
\label{eq:sigma_theta}
\end{equation}
We also define $\sigma_e$ as the RMS time-averaged eccentricity of all planets:
\begin{equation}
\sigma_{e} \equiv \left( \frac{1}{N} \langle \sum_{j} |e_j|^2 \rangle \right)^{1/2}.
\label{eq:sigma_e}
\end{equation}
It is also useful to consider an ``averaged'' coupling parameter, analogous to $\epsilon_{12}$ for the $N = 2$ case. We define the ``dominant'' planet (labeled ``d'') in the system as whichever planet in the system that has the largest mass. If all planets share the same mass, a good approximation is to let the planet with the median semi-major axis be the ``dominant'' one. \cite{Lai2017} found a good choice for an averaged coupling parameter $\bar{\epsilon}$ to be
\begin{equation}
\bar{\epsilon} \equiv \left( \frac{1}{N-1} \langle \sum_{j \neq d} |\epsilon_{jd}|^2 \rangle \right)^{1/2}.
\end{equation}


\subsection{Multi-planet Eccentricity Evolution : Linear Theory}
\label{sec:sec5.1}
For $e_p \ll 1$, it is suitable to use the Laplace-Lagrange theory. The evolution of the eccentricity vector of each planet is given by Eq. (\ref{eq:LL_e}), and we start by first casting this equation into matrix form:
\begin{equation}
\frac{d \mathbf{\mathcal{E}}}{dt} =  i \mathbf{A} \mathbf{\mathcal{E}}- i \mathbf{B} e_p,
\end{equation}
where $\mathcal{E}$ is an N-dimensional vector with element $\mathcal{E}_j$ given by the complex eccentricity (see Eq. \ref{eq:complex_e}). The matrix $\mathbf{A}$ is given by:
\begin{equation}
\mathbf{A} = 
 \begin{pmatrix}
  \omega_{1} & -\nu_{12} & \cdots & -\nu_{1N}  \\
  -\nu_{21} & \omega_{2} & \cdots & -\nu_{2N}   \\
  \vdots  & \vdots  & \ddots & \vdots \\
  -\nu_{N1} & -\nu_{N2} & \cdots & \omega_{N} 
 \end{pmatrix}
 \label{eq:A_matrix}
\end{equation}
with $\omega_j$ defined as the sum of ``quadrupole'' frequencies from all other planets acting on planet $j$, 
\begin{equation}
\omega_j \equiv \sum_{k \neq j}^{N} \omega_{jk} + \omega_{jp}.
\label{eq:omega_j}
\end{equation}
The vector $\mathbf{B}$ is an N-dimensional vector representing the forcing term given by ${B}_j = \nu_{jp}$ (where $j = 1, 2, 3... N$). 

Let $\mathbf{V}$ be the N$\times$N matrix of eigenvectors for $\mathbf{A}$, and $\lambda_n$ be the eigenvalue associated with the $n$-th eigenvector $\mathbf{V}_n$. Then the eccentricity evolution of the $j$-th planet is given by:
\begin{equation}
\mathcal{E}_j(t) = \mathcal{E}_{fj} + \sum_n^N b_n (\mathrm{V_n})_j \exp{(i\lambda_n t)} .
\label{eq:e_multi_t}
\end{equation}
where $(\mathrm{V_n})_j$ is the $j$-th component of vector $\mathrm{V_n}$, and $\mathcal{E}_{fj}$ is the forced eccentricity on the $j$-th planet given by 
\begin{equation}
\mathbf{\mathcal{E}}_{fj} = (\mathbf{A}^{-1} \mathbf{B})_j e_p.
\label{eq:e_f_multi}
\end{equation}
The co-efficient $b_n$ can be obtained by matching Eq. (\ref{eq:e_multi_t}) to the initial condition. Since the inner planet eccentricities are initially zero, we have that:
\begin{equation}
b_n = e_p \left(\mathbf{V}^{-1} \cdot \mathbf{A}^{-1} \cdot \mathbf{B}\right)_n.
\label{eq:bn}
\end{equation}
The RMS eccentricity of planet $j$ is then given by:
\begin{equation}
\langle e_j^2 \rangle^{1/2} = \left( \sum_n b_n^2 (\mathrm{V_n})_j^2  + \mathcal{E}_{fj}^2  \right)^{1/2}.
\label{eq:rms_e_gen}
\end{equation}
Analogous to the $N = 2$ case, in either the strong-coupling limit ($\bar{\epsilon} \ll 1$) or the weak-coupling limit ($\bar{\epsilon} \gg 1$), the forced eccentricity term dominates over the other modes, and a good approximation is
\begin{equation}
\langle e_j^2 \rangle^{1/2} = \sqrt{2} \mathcal{E}_{fj}.
\label{eq:rms_e_strong_coupling}
\end{equation}
The RMS eccentricity of the system $\sigma_e$ is then given by
\begin{equation}
\sigma_{e} \approx \sqrt{\frac{2}{N}}  |\mathbf{A^{-1}B}| e_p.
\label{eq:rms_e_sigma}
\end{equation}

Note that in the strong coupling limit, $\sigma_e$ and $\langle e_j^2 \rangle^{1/2}$ both scale proportionally to $m_p / a_p^4$, since in this limit $\omega_{jk} \gg \omega_{jp}$ for all $j, k \in \{1,2,...,N\}$, and $\mathbf{A^{-1}}$ is a linear combination of $\omega_{jk}$, while $\mathbf{B}$ is determined by $(\nu_{1p}, \nu_{2p}, ... \nu_{Np})$. Therefore, the quantity  $(\mathbf{A}^{-1} \mathbf{B})$ can be written as a vector whose entries are a linear combination of $\nu_{jp}$ divided by a linear combination of $\omega_{jk}$ (with various $j$ and $k$). Similarly, in the weak coupling limit we have $\omega_{j} \simeq \omega_{jp}$, and $\sigma_e$ and $\langle e_j^2 \rangle^{1/2}$ will scale with linear combinations of $\nu_{jp}/\omega_{jp}$ (with various $j$).

We illustrate our linear results on the top-left panel of Fig. \ref{fig:fig10}, where we plot $\sigma_e$ as a function of $\bar{\epsilon}$ (adjusted by adjusting $a_p$) for a 4-planet system under the influence of a $3M_J$ perturber with $e_p = \theta_p = 0.1$. The three different colored curves represent three different planet systems with different mass ratios between the dominant planet ($m_3 = m_d$) and the other planets (which have equal masses). The solid and dashed curves are computed from Eqs. (\ref{eq:rms_e_gen}) and (\ref{eq:rms_e_strong_coupling}) respectively, while the filled squares and `$\times$' markers are obtained from N-body integrations with squares and crosses representing stable and unstable systems. Notice the excellent agreement between the theoretical and numerical results.

\subsection{Multi-planet Inclination: Linear Theory}
\label{sec:sec5.2}
For $\theta_p \ll 1$, the evolution of the angular momentum vector of the $j$-th planet $\jvec_j$ can be obtained using Laplace-Lagrange theory (Eq. \ref{eq:LL_l}). Again we start by first re-writing this equation into matrix form:
\begin{equation}
\frac{d}{dt} \mathbf{\mathcal{I}} =  i \mathbf{C} \mathbf{\mathcal{I}} + i \mathbf{D} \theta_p.
\end{equation}
The matrix $\mathbf{C}$ is given by
\begin{equation}
\mathbf{C} = 
 \begin{pmatrix}
  -\omega_{1} & \omega_{12} & \cdots & \omega_{1N}  \\
  \omega_{21} & -\omega_{2} & \cdots & \omega_{2N}   \\
  \vdots  & \vdots  & \ddots & \vdots \\
  \omega_{N1} & \omega_{N2} & \cdots & -\omega_{N},
 \end{pmatrix}
 \label{eq:C_matrix}
\end{equation}
where $\omega_j$ is given by Eq. (\ref{eq:omega_j}) and the vector $\mathbf{D}$ is an N-dimensional vector representing the forcing term given by ${D}_j = \omega_{jp}$.

Let $\mathbf{Y}$ be the $N\times N$ matrix of eigenvectors for $\mathbf{C}$, and $\lambda_n$ be the eigenvalue associated with the $n$-th eigenvector $\mathbf{Y}_n$. Then the inclination evolution of the $j$th planet is given by:
\begin{equation}
\mathcal{I}_j(t) = \mathcal{I}_p + \sum_n^N c_n (\mathrm{Y}_n)_j \exp{(i\lambda_n t)}.
\end{equation}
The coefficients $c_n$ are determined by from the initial conditions, and are given by
\begin{equation}
c_n = \theta_p (\mathbf{Y}^{-1} \cdot \mathbf{C}^{-1} \cdot \mathbf{D})_n.
\label{eq:cn}
\end{equation}
The RMS mutual inclination between planet $j$ and $k$ is then given by
\begin{equation}
\langle \theta_{jk}^2 \rangle^{1/2} \simeq \sqrt{2} \sum_n^N c_n^2 [(\mathrm{Y}_n)_j - (\mathrm{Y}_n)_k]^2.
\label{eq:theta_p_pairwise}
\end{equation}
The RMS mutual inclination, RMS-averaged over all pairs of planets is given by:
\begin{equation}
\sigma_{\theta} \simeq \sqrt{2}  \left(\sum_n^N c_n^2 \Theta_{n}\right)^{1/2},
\label{eq:theta_p_multi}
\end{equation}
where $\mathbf{\Theta}$ is the N-dimensional vector given by:
\begin{equation}
\Theta_{n} = \sum_j^N (\mathrm{Y}_n)_j^2 - \frac{1}{N}\left(\sum_j^N (\mathrm{Y}_n)_j \right)^2.
\end{equation}
Note that in the strong coupling limit we have that $\omega_{jk} \gg \omega_{jp}$ for all $j, k \in \{1,2,...,N\}$, and therefore $\mathbf{C^{-1}}$ is given by some linear combinations of $\omega_{jk}$, while $\mathbf{D}$ is determined by $(\omega_{1p}, \omega_{2p}, ... \omega_{Np})$. Therefore, in this limit the RMS inclination scales as $\omega_{jp}/\omega_{jk}$ for various combinations of $j$ and $k$. On the other hand, in the weak coupling limit, we have $\omega_{j} \simeq \omega_{jp}$, and thus $\sigma_{\theta}$ is determined by by various combinations of ${\omega_{jp}}/{\omega_{kp}}$ (for various $j$ and $k$), and this combination approaches unity for $\bar{\epsilon} \gg 1$. 

The above results are compared against numerical integrations in the bottom-left panel of Fig. \ref{fig:fig10}, where we plot $\sigma_{\theta}$ as a function of $\bar{\epsilon}$ (adjusted by adjusting $a_p$) for a 4-planet system under the influence of a $3M_J$ perturber with $e_p = \theta_p = 0.1$. The three different colored curves represent three systems with different mass ratios between the dominant planet and the other planets (which have equal masses). The solid curves are computed from Eq. (\ref{eq:theta_p_multi}) while the filled squares and `$\times$' markers represent N-body integrations with stable and unstable systems respectively. Again, we find our analytical results to agree with numerical integrations.

\subsection{Extension to Modest Eccentricities and Inclinations}

The results in Sections \ref{sec:sec5.1} and \ref{sec:sec5.2} are derived under the assumption of $e_p$ and $\theta_p \ll 1$. For systems with modest values of $e_p$ and $\theta_p$ (up to $\sim 0.4$), we can extend the results to include the effect of finite $e_p$ and $\theta_p$. The derivation is analogous to the case of two inner planets as discussed in Section \ref{sec:secular_equations}. The results of Eqs. (\ref{eq:e_multi_t}) - (\ref{eq:rms_e_strong_coupling}) and (\ref{eq:theta_p_pairwise}) - (\ref{eq:theta_p_multi}) remain valid for modest values of $e_p$ and $\theta_p$ as long as one uses the modified frequencies $\tilde{\omega}_{jp}$ and $\tilde{\nu}_{jp}$ (see Eqs. \ref{eq:omega_tilde} - \ref{eq:nu_tilde}) instead of ${\omega_{jp}}$ and ${\nu_{jp}}$.

For a system of $N > 2$ inner planets in the strong coupling regime ($\bar{\epsilon} \ll 1$) the extension to modest $e_p$ and $\theta_p$ can be further simplified (analogous to Eqs. (\ref{eq:nonlin_extend_strong_a}) - (\ref{eq:nonlin_extend_strong_c}) for the $N = 2$ case):
\begin{align}
\langle         e_j^2 \rangle^{1/2}(e_p, \theta_p) &\simeq \langle e_j^2 \rangle^{1/2}_{\mathrm{lin}}  \left[\frac{1 - 5\theta_p^2/4}{(1 -e_p^2)^{5/2}}\right],
\label{eq:nonlin_extend_strong_multi_e} \\
\langle \theta_{jk}^2 \rangle^{1/2}(e_p, \theta_p) &\simeq  \langle \theta_{jk}^2 \rangle^{1/2}_{\mathrm{lin}}  \left[\frac{1 - \theta_p^2/2}{(1 -e_p^2)^{3/2}}\right],
\label{eq:nonlin_extend_strong_multi_th}
\end{align}
where $\langle e_j^2 \rangle^{1/2}_{\mathrm{lin}} $ and $\langle \theta_{jk}^2 \rangle^{1/2}_{\mathrm{lin}}$ are the expressions obtained from the linear theory, and are given by Eqs. (\ref{eq:rms_e_gen}) and (\ref{eq:theta_p_pairwise}) respectively. We omit the scaling for the weak coupling limit ($\bar{\epsilon} \gg 1$) here since weakly coupled $N > 2$ systems with modest $e_p$ are generally unstable.

The right panels of Fig. \ref{fig:fig10} show a comparison of the above analytical results with the results from numerical N-body integrations (the left panels show a case with $e_p = \theta_p = 0.1$ where the linear theory is approximately valid; the right panels show a set-up with more significant values of $e_p = \theta_p = 0.4$). We find that for the $e_p = \theta_p = 0.4$ case, our analytical results agreed well with the N-body simulations up to $\bar{\epsilon} \sim 1 $, at which point the system generally becomes unstable and the planets attain eccentricities much larger than predicted from the secular theory.

Combined with the results of Sections \ref{sec:sec5.1} and \ref{sec:sec5.2}, Eqs. (\ref{eq:nonlin_extend_strong_multi_e}) - (\ref{eq:nonlin_extend_strong_multi_th}) present a way to rapidly compute analytically the RMS planet eccentricities and mutual inclinations in a ``N-planets + perturber'' system without resorting to numerical integrations. A short summary for the steps to carry out this computation is given in Appendix \ref{sec:appendix2}. 

\begin{figure*}
\includegraphics[width=0.95\linewidth]{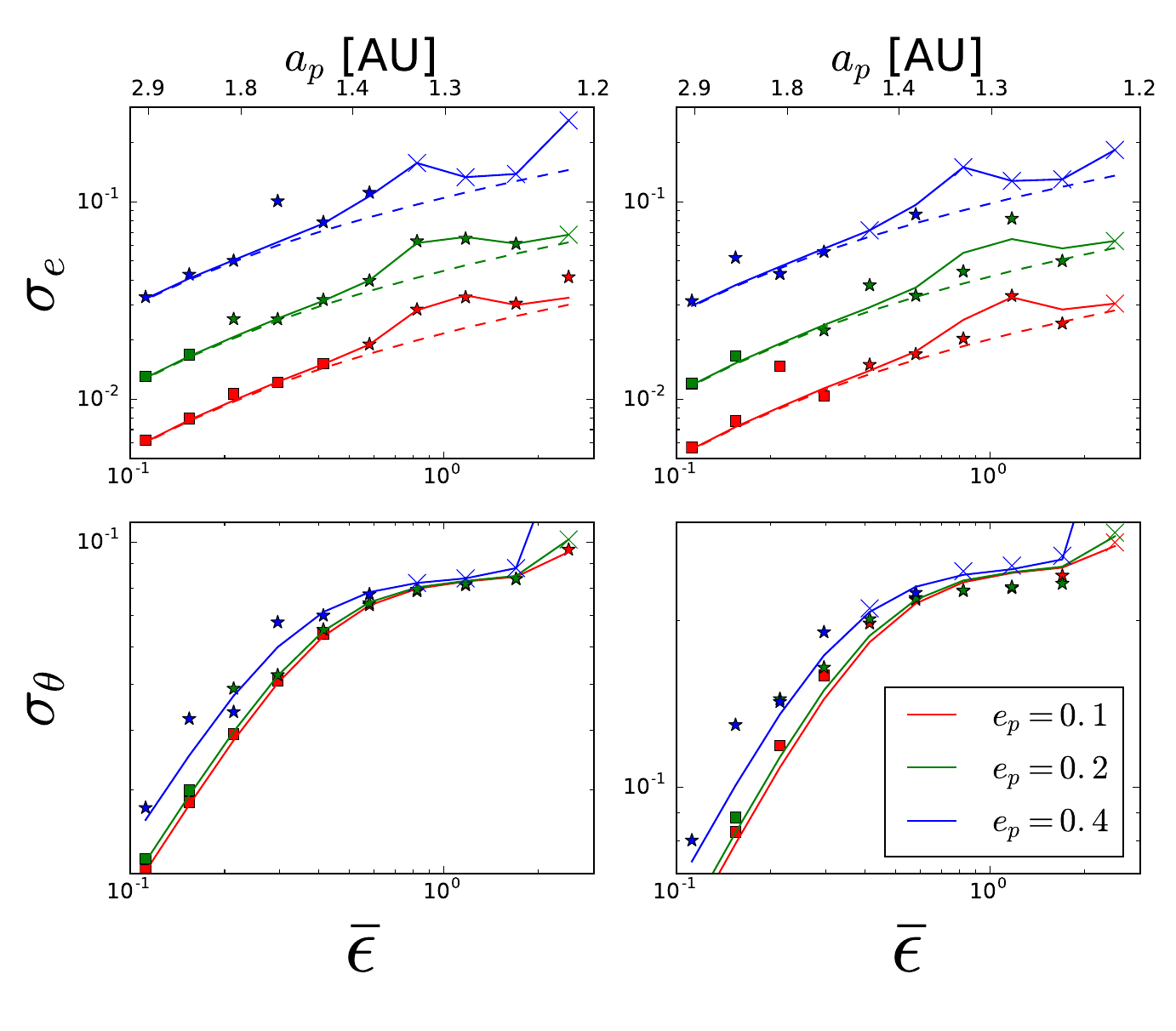}
\caption{RMS orbital eccentricity and mutual inclination (Eqs. \ref{eq:sigma_theta} - \ref{eq:sigma_e}) of the Kepler-11 system when being perturbed by a hypothetical misaligned and eccentric external companion. The top panels show the RMS eccentricity while the bottom panel shows the RMS mutual inclination. The perturber, whose coupling strength is parametrized by $\bar{\epsilon}$, has mass $3 M_J$ and semi-major axis ranging from 1 - 3 au; its eccentricity is  $e_p = 0.1$ for the left panels, and $e_p = 0.4$ for the right panels.  The inner 6 planets are initially started on circular, co-planar orbits. The red, green and blue curves correspond to different values of $\theta_p = $ 0.1, 0.2, 0.4 respectively. The solid lines are obtained from our hybrid secular theory (Eqs. \ref{eq:nonlin_extend_strong_multi_e} and \ref{eq:nonlin_extend_strong_multi_th} for the top and bottom panels respectively), while the points are obtained from N-body integrations over a period of $10^6$ yrs; for the top panels only, the dashed curves were obtained under the ``strong coupling'' approximation using Eq. (\ref{eq:rms_e_strong_coupling}). Filled squares represent systems stable against orbit crossings, while stars are systems that have undergone orbit crossings (but not collisions or ejections) within $10^6$ yrs. An `$\times$' marks an unstable systems where one or more planets have collided or been ejected. Systems marked by `$\times$' are arbitrarily placed on the solid curves for visual clarity as their RMS eccentricities and mutual inclinations can be ill-defined due to ejections and/or collisions.}
\label{fig:fig11}
\end{figure*}

\begin{figure*}
\includegraphics[width=0.95\linewidth]{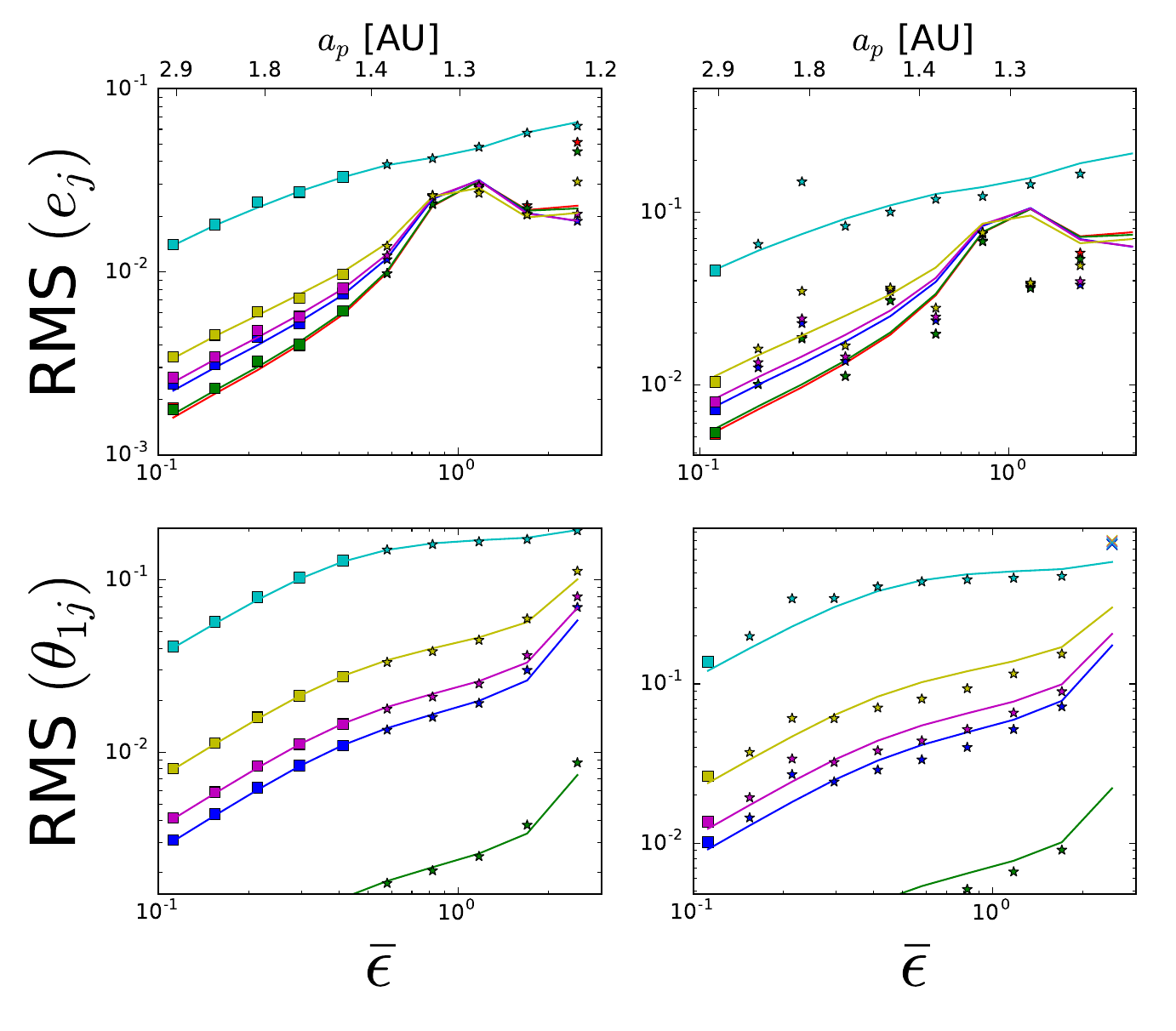}
\caption{The same as Fig. \ref{fig:fig11}, but showing the RMS orbital eccentricities $\langle e_j^2 \rangle^{1/2}$ (top panels) and mutual inclinations between the Kepler-11b and the $j$-th planet $\langle \theta_{1j} \rangle^{1/2}$ (bottom panel). The left panels have $e_p = \theta_p = 0.1$ while the right panels have $e_p = \theta_p = 0.3$. The red, green, blue, magenta, yellow and cyan curves correspond to planets $b$ - $g$ in the Kepler-11 system. The solid curves are obtained from the hybrid secular analytical theory, and given by Eqs. (\ref{eq:nonlin_extend_strong_multi_e}) and (\ref{eq:nonlin_extend_strong_multi_th}) for the top and bottom panels respectively.}
\label{fig:fig12}
\end{figure*}

\section{Application to the Kepler-11 System}
\label{sec:sec6}

In this section, we apply our results to Kepler-11, a system with 6 tightly super-Earths. Previous works have used the co-planarity of the 6 planets to constrain the presence of any misaligned external companions \citep{JontofHutter2017}. Due to the highly compact nature of this system, secular integrations were thought to be unsuitable and \cite{JontofHutter2017} relied on N-body simulations. Here we show that the secular theory based on our hybrid secular equations can robustly reproduce N-body orbital eccentricities and mutual inclinations of Kepler-11 under the influence of a companion for a wide range of parameter space.

We consider the Kepler-11 system \citep[with parameters as described in][]{Lissauer2011} with the addition of a $3 M_J$ companion ranging from 1.2 - 3 au. Planets $b$ through $g$ are given semi-major axes [0.091, 0.107, 0.155, 0.195, 0.250, 0.466] au, masses [1.9, 2.9, 7.3, 8.0, 2.0, 25] $M_{\oplus}$ and radii [1.8, 2.87, 3.12, 4.19, 2.49, 3.33] $R_{\oplus}$ respectively. Their initial orbital eccentricities and inclinations are set to zero while the perturber has its eccentricity and inclination set to range from $0.1$ to $0.4$. We integrate this system using the HERMES integrator from REBOUND (Rein et al 2015, Silburt \& Rein unpublished) instead of the WHFast integrator used in Sec. \ref{sec:sec4} as it offers superior accuracy over repeated close encounters between planets. The HERMES ``$R_{\mathrm{Hill}}$ switch factor'' is set to 1 and physical collisions between planets were assumed to be perfect mergers, an assumption that is reasonable for planets above $1 M_{\oplus}$ \citep[see][]{Mustill2017b}.

We then compute the RMS orbital eccentricities and mutual inclinations obtained from the N-body integrations and compare them to the secular analytical results described in Section 5 (shown in Figs. \ref{fig:fig11} and \ref{fig:fig12}). In Fig. \ref{fig:fig11}, the left and right panels represent two different inclinations for the perturber ($\theta_p = 0.1$ and $0.4$). The squares, stars and crosses represent systems that are stable, meta-stable and unstable respectively, with `meta-stable' referring to systems having undergone orbit crossings but not physical collisions and/or ejections with-in 1 Myr, while `unstable' refers to systems that have  undergone collisions and/or ejection events. We see that so long as the system is not `unstable', our hybrid secular theory shows excellent agreement with N-body simulations, even for mild values of $e_p = \theta_p = 0.4$. In Fig. \ref{fig:fig12}, we show the RMS values of $e_j$ and $\theta_{1j}$ (the mutual inclination between Kepler-11b and each of the other Kepler-11 planets) for all the individual planets of Kepler-11. The left and right panel represent two different perturber eccentricities and inclinations, with the left panel having $e_p = \theta_p = 0.1$ and the right panel having $e_p = \theta_p = 0.3$. For the case with $e_p = \theta_p = 0.1$ (left panels) we found our theoretical results to agree excellently with N-body simulations, while for the case with $e_p = \theta_p = 0.3$ our results agreed qualitatively with N-body simulations, with an average deviation of $\sim 30\%$ and a maximum deviation of $\sim 100\%$.

In summary, we find that a hypothetical, nearly co-planar $3 M_J$ perturber with $e_p = 0.1$ would drive the inner system into instability if it is closer than $a_p \approx 1.5$ au; the condition becomes $a_p \approx 2.3$ au if the perturber instead has $e_p = 0.2$.
\section{Summary and Discussion}
\label{sec:sec7}

We have studied the excitation of orbital eccentricities and mutual
inclinations in compact multi-planet systems induced by the
gravitational influence of eccentric and/or misaligned external
planetary or stellar companions.  Our major goal is to derive
anaytical expressions and scaling relations for the excited
eccentricities/inclinations of the inner system as functions of the
parameters the external perturber (mass $m_p$, semi-major axis $a_p$,
eccentricity $e_p$ and inclination $\theta_p$), so that the impact of
the perturber can be evaluated without resorting to computationally intensive N-body
integrations for a variety of systems systems.  We provide a summary of our main results and guide to key equations and figures as follows.

$\bullet$ For $e_p,\theta_p\ll 1$, we used the linear Laplace-Lagrange theory
to obtain explicit analytic expressions for the RMS mutual inclination $\theta_{12}$
(Eq.~\ref{eq:theta12}) and eccentricities $e_1,~e_2$
(Eqs.~\ref{eqn:gen_e1}-\ref{eqn:gen_e2}) of two inner planets
perturbed by an external companion (Section \ref{sec:sec2}).

In general, the dynamics of a ``2-planet + perturber'' system is
determined by the dimensionless parameter $\epsilon_{12}$ (Eq.\ref{eq:ep12}),
given by the ratio of the differential precession frequency of the
inner planets (driven by the perturber) and their mutual coupling
frequency (see also Lai \& Pu 2017). When the two inner planets are
weakly coupled ($\epsilon_{12} \gg 1$), they are more susceptible to
inclination and eccentricity excitations, with
\begin{align}
\theta_{12} &\sim\theta_p \\
e_j &\sim (a_j/a_p)e_p
\end{align}
(see Eqs.~\ref{eq:theta_12_weak}, \ref{eq:e1_weak} and \ref{eq:e2_weak}).
In contrast, a strongly coupled planet pair ($\epsilon_{12} \ll 1$)
experiences reduced inclination/eccentricity excitations, with
\begin{align}
\theta_{12} &\sim \epsilon_{12}\theta_p \propto (m_p/a_p^3) \theta_p \\
e_1,e_2 &\propto (m_p/a_p^4) e_p
\end{align}
(see Eqs.~\ref{eq:theta_12_strong}, \ref{eqn:strong_coupled_e1} and
\ref{eqn:strong_coupled_e2}).  This indicates that a pair of planets in
a compact configuration are more resistant to perturbations by a
misaligned and/or eccentric external companion, as compared to a more
loosely packed planet pair or a single planet. There may be observational support for this
trend: \cite{Xie2016} found that taken as a group, transiting Kepler
singles have systematically larger eccentricities than Kepler multis
(with $\bar{e} \sim 0.3$ for Kepler singles and 
$\bar{e} \sim 0.04$ for Kepler multis). Our results suggest
that perturbations by outer companions can be one contributing factor
to this observational trend.

For the case of $\epsilon_{12} \sim 1$, a resonance feature occurs if
the innermost planet is the least massive one (i.e. $m_1 \lesssim
m_2$), and the mutual inclinations and eccentricities of the inner planets can
be boosted to values much larger than $\theta_p$ and $e_p$
respectively (Eqs.~\ref{eqn:res_e1} - \ref{eqn:res_e2}).

$\bullet$ We extended our linear results to perturbers with
`modest' $e_p$ and $\theta_p$ by developing `hybrid' secular equations
of motion (Eqs.~\ref{eq:A1} - \ref{eq:A4}) that interpolates between
Laplace-Lagrange theory and multi-pole expansion (Section \ref{sec:sec3}).
We derived analytical results for the inner planet eccentricities
(Eqs.~\ref{eq:LL_e_nonlin}, \ref{eq:nonlin_extend_strong_a} -
\ref{eq:nonlin_extend_strong_b} and \ref{eq:nonlin_extend_weak_a} -
\ref{eq:nonlin_extend_weak_b}) and inclinations
(Eqs.~\ref{eq:LL_t_nonlin}, \ref{eq:nonlin_extend_strong_c} and
\ref{eq:nonlin_extend_weak_c}). 
Comparing with numerical integrations of hybrid secular equations and
N-body simulations (see Figs.~\ref{fig:fig4}-\ref{fig:fig6} and
Section \ref{sec:sec4}), we found that our analytical results are valid
for general values of $\theta_p$ and $e_p$ (up to $\theta_p, ~e_p \approx 0.4$), provided that the
resulting $e_j$ and $\theta_{12}$ are small.
In particular, for the cases where $\epsilon_{12} \sim 1$ and
$\theta_{12} \ge 0.68$ rad. ($39^{\circ}$), the inner planets can develop
Kozai-Lidov-like oscillations in their eccentricities and
inclinations, leading to rapid growth in $e_1$ and $e_2$ even for
small values of $e_p$, a feature that is not captured by our secular
theory. 

$\bullet$ 
We extended our analysis to inner systems with more than two
planets (Section \ref{sec:sec5}). In the linear theory, the inner
planet eccentricities are given by Eqs.~(\ref{eq:e_multi_t}) -
(\ref{eq:rms_e_strong_coupling}), and the mutual inclinations are
given by Eqs. (\ref{eq:theta_p_pairwise})-(\ref{eq:theta_p_multi}).
For strongly coupled inner systems, we extend our linear results to
modest values of $e_p$ and $\theta_p$ in
Eqs.~(\ref{eq:nonlin_extend_strong_multi_e}) and
(\ref{eq:nonlin_extend_strong_multi_e}). A comparison of these results
are shown against N-body simulations for a hypothetical ``4-planet +
perturber'' system in Fig. \ref{fig:fig9}. The results of our linear
theory agree robustly with N-body simulations, as long as the inner
system is not made unstable by the external perturber.

$\bullet$ We applied our hybrid secular equations to Kepler-11, a tightly packed
6-planet system. We examined the impact of a hypothetical external giant planet 
companion on the observed system,
and compared the results of `hybrid' secular equation integrations
with N-body simulations, the results of which are shown in
Figs.~\ref{fig:fig10} and \ref{fig:fig11}. We found our hybrid
secular theory to agree closely with N-body simulations as long as the
Kepler-11 system is not rendered unstable against collisions and/or
ejections within the integration timescale.  For example, using on
hybrid secular equations, we can rule out the presence of a $\sim 3
M_J$ companion to Kepler-11 with eccentricity $e_p = 0.1$ out to $a_p
= 1.4$ au, and $e_p = 0.2$ out to $a_p = 2.5$ au.

In this work we have focused on evolution of an inner multi-planet
system with initially circular, co-planar orbits subject to an
eccentric and/or misaligned external companion. How might such a
companion be generated is a pertinent question that lies outside the
scope of this work. In the case of distant stellar companions, the
eccentricity and inclination of the perturber is simply a product of
star/binary formation process in a turbulent molecular cloud.  In the
case of giant planet companions, the eccentricity and inclination may
be generated as the end-product of a violent scattering process in an
unstable system of primordial giant planets that underwent violent close
encounters and scatterings until only a single planet remained.
In this scenario, the assumption that such inner systems
have initially circular and co-planar is likely to be incorrect, as the
violent process itself may generate inner planet eccentricities and
mutual inclinations larger than than the results found in this
work. Nevertheless, the secular results studied in this paper provide a 
benchmark of the eccentricity/inclination excitation by the external companions.
In an upcoming paper (Pu \& Lai 2018 in prep) we will study the
scenario involving primordial giant planet scatterings, and present a
model for the inner planet eccentricity and mutual
inclinations excitation during the violent giant-planet scattering phase.

\section*{Acknowledgements}
This work has been supported in part by NASA grants
NNX14AG94G and NNX14AP31G, and NSF grant AST-1715246.
BP thanks NASA for the NESSF fellowship and the Canadian Institute for
Theoretical Astrophysics for computing resources.

\bibliography{msNotes}
\bibliographystyle{mnras}

\appendix

\section{Hybrid Secular Equations}
\label{sec:appendix}

Our hybrid secular equations are based on the equations given by \cite{Liu2015} that govern the secular evolution of
hierarchical triples (where the semi-major axes of the inner and outer
binaries satisfy $a_{\rm in}\ll a_{\rm out}$) with arbitrary
eccentricities and inclinations.  These equations are expressed in
terms of the dimensionless angular momentum vector and eccentricity vector,
\begin{equation}
{\bf j}=\sqrt{1-e^2}{\bf \hat n},\quad 
{\bf e}=e\,{\bf\hat u}
\end{equation}
(where ${\nvec}$ and ${\uvec}$ are unit vectors),
and extend previous results \citep[e.g.][]{Milankovic1939, Tremaine2009}
by expanding the interaction potential to the octupole order \citep[see also][]{Boue2014, Petrovich2015_Secular}.
While the \cite{Liu2015} equations accurately capture the interaction between a planet
in the inner system and the distant perturber, they are not valid for describing the
interaction between the inner planets. We therefore modify these equations
by replacing the quadrupole and octupole strengths with ones given
by appropriate Laplace coefficients in the standard Laplace-Lagrange secular theory.
Obviously, the Laplace-Lagrange theory is valid only for $e_j, ~\theta_j\ll 1$.
But when the inner planets develop large eccentricities and/or mutual inclinations,
dynamical instability is likely to set in.  


In our hybrid equations, the rates of change of the dimensionless angular
momentum vector $\jvec_j$ and eccentricity vector $\evec_j$ of an inner planet
$j$ induced by an outer planet $k$ (including the perturber planet $p$) are given by:
\begin{equation}
\begin{split}
\left(\frac{d{\jvec_{j}}}{dt}\right)_k &= \frac{\omega_{jk}}{(1-e_k^2)^{3/2}} \Big[(\jvec_j\cdot\nvec_k)~\jvec_j\times\nvec_k
-5(\evec_j\cdot\nvec_k)~\evec_j\times\nvec_k\Big]\\
&-\frac{5\nu_{jk}e_k}{4(1-e_k^2)^{5/2}}\Bigg\{
\bigg[2\Big[(\evec_j\cdot\uvec_k)(\jvec_j\cdot\nvec_k)\\
&+(\evec_j\cdot\nvec_k)(\jvec_j\cdot\uvec_k)\Big]~\jvec_j+2\Big[(\jvec_j\cdot\uvec_k)(\jvec_j\cdot\nvec_k)\\
&-7(\evec_j\cdot\uvec_k)(\evec_j\cdot\nvec_k)\Big]~\evec_j\bigg]\times\nvec_k\\
&+\bigg[2(\evec_j\cdot\nvec_k)(\jvec_j\cdot\nvec_k)~\jvec_j
+\Big[\frac{8}{5}e_j^2-\frac{1}{5}\\
&-7(\evec_j\cdot\nvec_k)^2+(\jvec_j\cdot\nvec_k)^2\Big]~\evec_j\bigg]
\times\uvec_k\Bigg\},
\end{split}
\label{eq:A1}
\end{equation}

\begin{equation}
\begin{split}
\left(\frac{d{\evec_{j}}}{dt}\right)_k &= \frac{\omega_{jk}}{(1-e_k^2)^{3/2}} \Big[(\jvec_j\cdot\nvec_k)~\evec_j\times\nvec_k
+2~\jvec_j\times\evec_j\\
&-5(\evec_j\cdot\nvec_k)\jvec_j\times\nvec_k\Big]\\
&-\frac{5\nu_{jk}e_k}{4(1-e_k^2)^{5/2}}\Bigg\{
\bigg[2(\evec_j\cdot\nvec_k)(\jvec_j\cdot\nvec_k)~\evec_j\\
&+\Big[\frac{8}{5}e_1^2-\frac{1}{5}-7(\evec_j\cdot\nvec_k)^2+(\jvec_j\cdot\nvec_k)^2\Big]~\jvec_j\bigg]\times\uvec_k\\
&+\bigg[2\Big[(\evec_j\cdot\uvec_k)(\jvec_j\cdot\nvec_k)+(\evec_j\cdot\nvec_k)(\jvec_j\cdot\uvec_k)\Big]~\evec_j\\
&+2\Big[(\jvec_j\cdot\nvec_k)(\jvec_j\cdot\uvec_k)-7(\evec_j\cdot\nvec_k)(\evec_j\cdot\uvec_k)\Big]~\jvec_j
\bigg]\times\nvec_k\\
&+\frac{16}{5}(\evec_j\cdot\uvec_k)~\jvec_j\times\evec_j\Bigg\}.
\end{split}
\label{eq:A2}
\end{equation}

Meanwhile, the outer planet $k$ being influenced by the inner planet $j$ is described by the equations:
\begin{equation}
\begin{split}
\left(\frac{d{\jvec_{k}}}{dt}\right)_j&= \frac{\omega_{kj}}{(1-e_k^2)^{3/2}} \Big[(\jvec_j\cdot\nvec_k)~\nvec_k\times\jvec_j
-5(\evec_j\cdot\nvec_k)~\nvec_k\times\evec_j\Big]\\
&-\frac{5\nu_{kj}e_k}{4(1-e_k^2)^{5/2}} \Bigg\{
2\Big[(\evec_j\cdot\nvec_k)(\jvec_j\cdot\uvec_k)~\nvec_k\\
&+(\evec_j\cdot\uvec_k)(\jvec_j\cdot\nvec_k)~\nvec_k+(\evec_j\cdot\nvec_k)(\jvec_j\cdot\nvec_k)~\uvec_k\Big]\times\jvec_j\\
&+\bigg[2(\jvec_j\cdot\uvec_k)(\jvec_j\cdot\nvec_k)~\nvec_k-
14(\evec_j\cdot\uvec_k)(\evec_j\cdot\nvec_k)~\nvec_k\\
&+\Big[\frac{8}{5}e_j^2-\frac{1}{5}-7(\evec_j\cdot\nvec_k)^2+(\jvec_j\cdot\nvec_k)^2\Big]~\uvec_k\frac{}{}\bigg]\times\evec_j
\Bigg\},
\end{split}
\label{eq:A3}
\end{equation}

\begin{equation}
\begin{split}
\left(\frac{d{\evec_{k}}}{dt}\right)_j&= \frac{\omega_{kj}}{(1-e_k^2)^{3/2}}
\bigg[
(\jvec_j\cdot\nvec_k)~\evec_k\times\jvec_j-5(\evec_j\cdot\nvec_k)\evec_k\times\evec_j\\
&-\Big[\frac{1}{2}-3e_1^2+\frac{25}{2}(\evec_j\cdot\nvec_k)^2-\frac{5}{2}(\jvec_j\cdot\nvec_k)^2\Big]\nvec_k\times\evec_k\bigg]\\
&-\frac{5\nu_{kj}e_k}{4(1-e_k^2)^{5/2}}\frac{e_2}{\sqrt{1-e_2^2}}\frac{L_1}{L_2}
\Bigg\{2\Big[(\evec_j\cdot\nvec_k)(\jvec_j\cdot\evec_k)~\uvec_k\\
&+(\jvec_j\cdot\nvec_k)(\evec_j\cdot\evec_k)~\uvec_k+\frac{1-e_2^2}{e_2}(\evec_j\cdot\nvec_k)
(\jvec_j\cdot\nvec_k)~\nvec_k\Big]\times\jvec_j\\
&+\bigg[2(\jvec_j\cdot\evec_k)(\jvec_j\cdot\nvec_k)~\uvec_k
-14(\evec_j\cdot\evec_k)(\evec_j\cdot\nvec_k)~\uvec_k\\
&+\frac{1-e_2^2}{e_2}\Big[\frac{8}{5}e_1^2-\frac{1}{5}-7(\evec_j\cdot\nvec_k)^2+(\jvec_j\cdot\nvec_k)^2\Big]~\nvec_k\bigg]
\times\evec_j\\
&-\bigg[2\left(\frac{1}{5}-\frac{8}{5}e_1^2\right)(\evec_j\cdot\uvec_k)~\evec_k\\
&+14(\evec_j\cdot\nvec_k)(\jvec_j\cdot\uvec_k)(\jvec_j\cdot\nvec_k)~\evec_k
+7(\evec_j\cdot\uvec_k)\Big[\frac{8}{5}e_1^2\\
&-\frac{1}{5}-7(\evec_j\cdot\nvec_k)^2+(\jvec_j\cdot\nvec_k)^2\Big]~\evec_k
\bigg]\times\nvec_k\Bigg\}.
\end{split}
\label{eq:A4}
\end{equation}

In the above equations, $L_j \simeq m_j \sqrt{GM_* a_j}$ is the angular momentum, and the quantities 
$\omega_{jk}$ and $\nu_{jk}$ are given by Eqs. (\ref{eq:wjk}) and (\ref{eq:vjk}) respectively.

For planet $j$, one would sum over the contributions from all other
planets according to the above formulae. Note that $j,k$ includes the perturber $p$. 
The time evolution of the $j$-th planet is thus:
\begin{align}
\frac{d{\jvec_{j}}}{dt} &= \sum_{k \neq j} \left(\frac{d{\jvec_{j}}}{dt}\right)_k,\\
\frac{d{\evec_{j}}}{dt} &= \sum_{k \neq j} \left(\frac{d{\evec_{j}}}{dt}\right)_k.
\end{align}
For $a_j\ll a_k$, we have
\begin{equation}
\omega_{jk} \simeq \frac{3Gm_j m_k a_j^2}{4a_k^3 L_j} ~ ~ ~ ~ ~ ~ ~ \mathrm{and} ~ ~ ~ ~ ~ ~
\nu_{jk} \simeq \frac{15Gm_j m_k a_j^3}{4a_k^4 L_j},
\end{equation}
and equations (\ref{eq:A1})-(\ref{eq:A4}) reduce to the equations (17)-(20) of \citep{Liu2015}.
For $e_j,e_k\ll 1$ and $\nvec_j\simeq \nvec_k$ (i.e., the mutual inclination between planets is small), equations (\ref{eq:A1})-(\ref{eq:A4}) reduce to the linearized Laplace-Lagrange equations given in Section \ref{sec:sec2}.

\section{A prescription for the eccentricities and mutual inclinations of ``N planets + perturber'' systems}
\label{sec:appendix2}
We summarize the short sequence of calculations that should be applied to determine the predicted RMS eccentricities and mutual inclinations and its regime of validity for a ``N planets + perturber'' system with inner planets on initially circular and co-planar orbits based on our hybrid secular equations. The necessary parameters required are the planet semi-major axes $a_j$, masses $m_j$ (with $j \in [1, 2, 3..., N, p]$) and the perturber's inclination and eccentricity $\theta_p$ and $e_p$. 

\begin{enumerate}
\item First, calculate the ``quadrupole'' and ``octupole'' precession frequencies $\omega_{jk}$ and $\nu_{jk}$ for all possible pairs of planets (including perturber $p$) from Eqs. (\ref{eq:wjk}) - (\ref{eq:vjk}).

\item For the perturber $p$ only, calculate the `adjusted' precession frequencies $\tilde{\omega}_{jp}$ and $\tilde{\nu}_{jp}$ for each of the inner planets $j \in [1, 2, 3..., N]$ (Eqs. \ref{eq:omega_tilde} - \ref{eq:nu_tilde}). From here onwards, all calculations involving the quantities $\omega_{jk}$ and $\nu_{jk}$ should be replaced with the tilded versions $\tilde{\omega}_{jp}$ and $\tilde{\nu}_{jp}$.

\item Compute the coupling matrices $\mathbf{A}$ and $\mathbf{C}$ from Eqs. (\ref{eq:A_matrix}) and (\ref{eq:C_matrix}). Note that $\tilde{\omega}_{jp}$ and $\tilde{\nu}_{jp}$ should be used in place of $\omega_{jk}$ and $\nu_{jk}$. 

\item Evaluate the $N \times N$ matrix of eigenvectors $\mathbf{Y}$ and $\mathbf{V}$ for the matrices $\mathbf{A}$ and $\mathbf{C}$ respectively.

\item Write down the forcing vectors $\mathbf{B}$ and $\mathbf{D}$, given by $\mathrm{B}_j = \tilde{\nu}_{jp}$ and $\mathrm{D}_j = \tilde{\omega}_{jp}$.

\item Using Eqs. (\ref{eq:bn}) and (\ref{eq:cn}), compute the co-efficients $b_n$ and $c_n$. 

\item The RMS eccentricity of the $j$-th planet $\langle e_j^2 \rangle^{1/2}$ is given by Eq. (\ref{eq:rms_e_gen}), and the RMS eccentricity of the system $\sigma_e$ is given by Eq. (\ref{eq:sigma_e}).

\item The RMS mutual inclination between the $j$-th and $k$-th planet $\langle \theta_{jk}^2 \rangle^{1/2}$ is given by Eq. (\ref{eq:theta_p_pairwise}), and the RMS mutual inclination of the system $\sigma_{\theta}$ is given by Eq. (\ref{eq:theta_p_multi}).

\item Check the planet pairs for mutual inclinations exceeding the Kozai critical angle: If any planet pairs have $(\theta_{jk})_{\mathrm{max}} \gtrsim 39^{\circ}$, they will under-go Lidov-Kozai-like oscillations and the hybrid secular equations break down, and one should resort to N-body integrations. The maximum mutual inclination is given approximately by $(\theta_{jk})_{\mathrm{max}} \simeq \sqrt{2} \langle \theta_{jk}^2 \rangle^{1/2}$. 

\item Check the planet pairs for orbital crossings: If any planet pairs have $a_j[1 + (e_j){\mathrm{max}}] \ge a_{j+1}[1 - (e_{j+1}){\mathrm{max}}]$, then their orbits cross and the hybrid secular equations break down; such systems are unstable and should be evaluated using N-body integrations. The maximum eccentricity is given approximately by $(e_j)_{\mathrm{max}} \simeq \sqrt{2} \langle e_j \rangle^{1/2}$.

\end{enumerate}

\end{document}